\documentclass[12pt,a4paper]{article}
\usepackage{amsmath}
\usepackage{longtable,threeparttable,booktabs}
\usepackage[small]{caption}

\usepackage[top=1.2in, bottom=1.2in, left=1.2in, right=1.2in]{geometry}

\usepackage[round]{natbib}
\usepackage{color,soul}
\usepackage{multirow}

\DeclareCaptionStyle{italic}[justification=centering]{labelfont={bf},textfont={it},labelsep=colon}
\usepackage{graphicx,psfrag,epsf,textcomp}
\usepackage{dsfont}
\usepackage{natbib,setspace}
\usepackage{url}
\usepackage[table,xcdraw]{xcolor}

\usepackage{booktabs, subfig, bm, paralist,mathpazo,tikz,longtable,microtype}
\usepackage{enumitem}

\usepackage[pdftex,colorlinks=true]{hyperref}
\definecolor{darkblue}{rgb}{0,0,.6}
\hypersetup{citecolor=darkblue,linkcolor=darkblue,urlcolor=darkblue}

\usepackage{csquotes}

\newcommand{\blind}{0}

\graphicspath{{plots/}}

\newsavebox\CBox

\usepackage[title]{appendix}
\usepackage{breqn}
\usepackage{bm,orcidlink}

\begin{document}

\def\spacingset#1{\renewcommand{\baselinestretch}{#1}\small\normalsize} \spacingset{1}

\if0\blind
{
  \title{\bf A model sufficiency test using permutation entropy}
  \author{
    Xin Huang \orcidlink{0000-0003-0176-1197} \qquad Han Lin Shang \orcidlink{0000-0003-1769-6430} \footnote{Corresponding author: Department of Actuarial Studies and Business Analytics, Level 7, 4 Eastern Road, Macquarie University, Sydney, NSW 2109, Australia; Telephone: +61(2) 9850 4689; Email: hanlin.shang@mq.edu.au}\\
    Department of Actuarial Studies and Business Analytics \\
    Macquarie University \\
    \\
	David Pitt \orcidlink{0000-0002-6976-7168} \\
Department of Economics \\
University of Melbourne	
  }
\maketitle
} \fi

\if1\blind
{
\title{\bf A model sufficiency test using permutation entropy}
\maketitle
} \fi

\bigskip

\begin{abstract}

Using the ordinal pattern concept in permutation entropy, we propose a model sufficiency test to study a given model's point prediction accuracy. Compared to some classical model sufficiency tests, such as the \citeauthor{broock1996test}'s \citeyearpar{broock1996test} test, our proposal does not require a sufficient model to eliminate all structures exhibited in the estimated residuals. When the innovations in the investigated data's underlying dynamics show a certain structure, such as higher-moment serial dependence, the \citeauthor{broock1996test}'s \citeyearpar{broock1996test} test can lead to erroneous conclusions about the sufficiency of point predictors. Due to the structured innovations, inconsistency between the model sufficiency tests and prediction accuracy criteria can occur. Our proposal fills in this incoherence between model and prediction evaluation approaches and remains valid when the underlying process has non-white additive innovation.

\vspace{.1in}

\end{abstract}

\spacingset{1.48}

\newpage

\section{Introduction}

The development of new tests used to evaluate forecasting models has remained an important subject over the last few decades. As an essential step in empirical time series analysis, practitioners need to know how well the employed models capture and replicate observed data dynamics. More importantly, model diagnostics consider how closely the constructed forecast predicts future movements. Several criteria and methods have been proposed to assess the performance of various prediction models. The commonly used model evaluation methods can be broadly categorized into four streams according to their rationale and objectives.

The first and most-used stream of model evaluation approaches is based on statistics reflecting the departure of the predicted values generated from a model from the actual realizations, sometimes with an additional term for parsimonious control. Typical examples include the R-square, Akaike Information Criterion, Bayesian Information Criterion, Mean Squared Error (MSE), Mean Absolute Error, Mean Absolute Percentage Error, and the Theil Inequality Coefficient \citep{leuthold1975use}. The former three measures are commonly computed on the in-sample data set. In contrast, the latter four are often computed on out-of-sample data that are not used to estimate a given model.

The second stream of approaches is based on predictive accuracy tests. A typical predictive accuracy test requires a pre-specified loss function, and the null hypothesis is usually set to be two or more competing models that have equal loss differential. Examples of predictive accuracy tests include \citeauthor{diebold2002comparing}'s \citeyearpar{diebold2002comparing} test, \citeauthor{clark2001tests}'s \citeyearpar{clark2001tests} test, \citeauthor{corradi2005test}'s \citeyearpar{corradi2005test} test and \citeauthor{hansen2011model}'s \citeyearpar{hansen2011model} model confidence set (MCS). The Diebold and Mariano (DM) test is used to compare two non-nested models. The Clark and McCracken (CM) test is designed to compare two nested models. The Corradi and Swanson (CS) test and MCS are designed to compare more than two models.

The third type of approach is based on economic criteria that use a simple trading strategy guided by forecasts to aggregate relative pay-offs generated by different forecasting models \citep[see, e.g.,][for more details]{granger2000economic,elliott2008economic}.
For instance, a simple investment rule can be constructed based on the out-of-sample one-period-ahead return forecast. For an initial endowment of a fixed monetary unit, an investor has to decide whether to maintain this wealth in cash or assets, depending on whether the return predicted for the next period exceeds a threshold given by the transaction cost. In the end, the final portfolio value will be compared between competing models. 

The last approach to model evaluation methods does not involve comparing the models under consideration. Instead, tests are conducted on the residuals to search for remaining structure to indicate the models' sufficiency. The most popular test used for residual analysis is \citeauthor{broock1996test}'s \citeyearpar{broock1996test} (BDS) test. The BDS test is the most often used test on residuals to check whether the residuals are independent; thereby, the estimated model is sufficient. The BDS test is a portmanteau test for measuring serial dependence. It can be used for testing against a variety of possible deviations from independence, including linear and non-linear dependence or chaos. 

Every model evaluation approach has its advantages and disadvantages. There is no universally accepted criterion for evaluating a model. The former three categories of methods are all sensible comparison criteria to differentiate model performance. However, they are only informative when used in comparison studies. They cannot reflect a given model's performance relative to the considered data's maximum prediction potential under investigation. The last approach, the independence test on residuals, can indicate the sufficiency of individual models. However, it suffers a major limitation in evaluating point forecasts as it presumes that a sufficient model is expected to generate independent innovations. For a series of innovations that exhibit higher-moment dependent structures, the residuals formed by the difference between the actual realizations and their oracle point predictors can violate white noise.

A typical example of a process with non-white additive innovations is the GARCH model. Suppose we consider a time series of the squared returns, the additive innovations of the GARCH model are not independent and identically distributed (iid) due to the heteroskedasticity present in the variance. In practice, when evaluating the sufficiency of the GARCH model, the BDS test is often applied to the standardized residuals $Y_{t}/\widehat{\sigma}_{t}$. However, an earlier study of \cite{brock1991nonlinear} suggests when the assumed model is not in a linear additive form, adjustment is required to ensure the BDS statistics have the right size under the null hypothesis. A straightforward approach is to apply the test on the natural logarithms of squared standardized residuals $\log(Y_{t}^{2}/\widehat{\sigma}_{t}^{2})$ so that the logarithmic transformation casts the GARCH model into a linear additive model \citep{fernandes2012finite}. However, the logarithm transformation may distort the structures left in standardized residuals and mislead the BDS test to accept an insufficient model incorrectly. In the appendix, we will provide a simulation study that shows how and why the BDS test on the natural logarithm transformed squared standardized residuals fails to reject an inadequate predictor/model.  

The GARCH model is not the only process that has non-white additive innovations. When the process governing the target time series exhibits higher-moment serial dependence, the case of non-white additive innovations can easily occur. 

A stationary time series $\{x_{t}:t=1,2,...,N\}$ can be represented or transformed into the form 
\begin{equation} \label{eqn:eqnTimeSeriesGForm}
{x}_{t}=c+g(x_{t-1},x_{t-2},...,x_{t-\tau_{T}})+\varepsilon_{t},
\end{equation} 
where $\tau_{T}$ is the furthest lag on which past entries affects the current entry $x_{t}$, $g(\cdot)$ is a deterministic function that connects past observations to the expected value of the current entry, $\text{E}(\varepsilon_{t})=0$ and $\text{E}(\varepsilon_{t}|F_{t-1})=0$ so that innovations cannot be further exploited to improve the forecasting accuracy in conditional mean of the current entry and $F_{t-1}$ denotes the information set available at time $t-1$. The representative additive form of time series in~\eqref{eqn:eqnTimeSeriesGForm} separates the first-moment structures from the higher-moment structures exhibited in the underlying dynamics. More specifically, the deterministic function $g(\cdot)$ determines the maximum potential of the point prediction capacity of the underlying process. Accordingly, the oracle one-step-ahead point predictions of the future variable $x_{t}$ is 
\begin{align*}
\widehat{x}^{\text{oracle}}_{t}=\text{E}(x_{t}|F_{t-1})=c+g(x_{t-1},x_{t-2},...,x_{t-\tau_{T}}).
\end{align*}
The additive innovation term $\varepsilon_{t}$ can deviate from white noise when there exist higher-order dependence structures that do not have impacts on $\text{E}(x_{t}|F_{t-1})$ but affect the higher-moment of the distribution of the response. 

To evaluate whether a point forecast $\widehat{x}_{t}$ is sufficiently close to the oracle prediction $\widehat{x}^{\text{oracle}}_{t}$, we develop a new model sufficiency test by inheriting the ordinal pattern concept used in a novel measure, Permutation Entropy (PE). Our test requires minimal prior knowledge or assumptions about the data generating process of the observed data or the model employed to make predictions and is specially designed to remain valid when the innovations $\varepsilon_{t}$ defined in~\eqref{eqn:eqnTimeSeriesGForm} violate white noise. 

PE is a complexity and predictability measure proposed by \cite{bandt2002permutation} based on ordinal patterns. Previous literature documented that the usage of ordinal patterns provides PE with many desirable properties. The primary advantages of PE reside in its flexibility to account for any form of structure, minimal requirements for prior assumptions and knowledge, robustness to observational and dynamic noise \citep{bandt2002permutation}. PE can deal with highly non-linear complex dynamics \citep{zunino2010permutation}, invariance to monotonic transformations, and robustness to non-stationarity \citep{kreuzer2014non}. 

In empirical applications, PE has been shown to be a powerful tool in capturing the characteristics of complex systems, such as brain (electrical) activity \citep{costa2002multiscale}, heart rate rhythms \citep{rubinov2010complex} and financial markets \citep{zunino2009forbidden}. The bivariate PE has been extended to measure the dependence structures of multivariate time series \citep{matilla2014permutation}. The successful PE applications in various disciplines and areas emphasize the ordinal-pattern-based measures' capability in reflecting and characterizing fundamental features of the dynamics governing the observed data. Using the ordinal patterns' predictability as a symbolic representation of the level of dependence, the proposed test demonstrates a strong power of discriminating sufficient and suboptimal point forecasts.  

The paper is organized as follows. Section~\ref{sec:2} provides the rationale and the specifications of the new point prediction sufficiency test. In the simulation studies of Section~\ref{sec:3}, we apply the proposed test to several generated time series governed by different forms of processes. Section~\ref{sec:4} illustrates the empirical application of our test on EUR/USD one-hour aggregated volatility series. In Section~\ref{sec:5}, we give the conclusions and final remarks.

\section{Rationale and specification of the PE model sufficiency test} \label{sec:2}

Our proposed test is built on comparing the level of dependence between residuals to that between residual and lagged observations of the prediction object. The relative relation of the quantities in the comparison is informative in determining whether a point forecast model is sufficient. 

Given two stationary univariate time series $X=\{x_{t}; t=1,2,\ldots,N\}$ and $Y=\{y_{t}; t=1,2,\ldots,N\}$, we create a new measure $\text{K}[(x_{t-D+2},x_{t-D+3},\dots,x_{t}),y_{t+\tau}]$ to quantify the strength of dependence between the historical observations in $X$ and a $\tau$-ahead future entry in $Y$. Let $\bm{x}^{D-1}_{t}$ denote a vector consisting $D-1$ number of consecutive lagged variables in $X$, i.e.
\begin{equation*}
\bm{x}^{D-1}_{t}=(x_{t-D+2},x_{t-D+3},\dots,x_{t}).
\end{equation*} 
The extent that $\bm{x}^{D-1}_{t}$ affects $y_{t+\tau}$ can be quantified through  
\begin{equation} \label{Ktheoretical}
\text{Dependence}(\bm{x}^{D-1}_{t},y_{t+\tau})\equiv \int_{y}\int_{x} \left[p_{(\bm{x}^{D-1}_{t},y_{t+\tau})}(x,y)-p_{\bm{x}^{D-1}_{t}}(x)p_{Y}(y)\right]^2 dxdy,
\end{equation} 
where $p_{(\bm{x}^{D-1}_{t},y_{t+\tau})}$ is the joint probability distribution between $\bm{x}^{D-1}_{t}$ and $y_{t+\tau}$, $p_{\bm{x}^{D-1}_{t}}$ is the marginal probability distribution of $\bm{x}^{D-1}_{t}$ and $p_{Y}$ is the marginal probability distribution of $Y$. By dependence we mean the more general definition of dependence that is not limited to linear dependence like the correlation coefficient. More specifically, the overall level of dependence between two variables is measured by the disparity between their joint distribution and the product of their respective marginal distributions.   

The value of the expression given in~\eqref{Ktheoretical} is hard to obtain, especially in empirical analysis where the underlying processes of $X$ and $Y$ are unknown. Therefore, we consider a way to approximate the expression given in~\eqref{Ktheoretical} by using the ordinal pattern. 

We first construct a collection of segments of length $D$ combining the observations in $\bm{x}^{D-1}_{t}$ and $y_{t+\tau}$ in the same vector, i.e.
\begin{equation*}
s_{t,\tau }^{D}=\left(x_{t-D+2}-\overline{x},\ldots,x_{t-1}-\overline{x},x_{t}-\overline{x},y_{t+\tau}-\overline{y}\right),  \qquad  t=D+\tau-1,D+\tau,\ldots,N-\tau,
\end{equation*}
where $\overline{x}$ and $\overline{y}$ represent the mean of the time series $\{x_{t};t=1,2,\ldots,N\}$ and $\{y_{t};t=1,2,\ldots,N\}$. The reason we subtract the mean of the time series in the constructed segments is to ensure the comparability of the entries from the two distinct series. Next we replace the last entry $y_{t+\tau
}-\overline{y}$ in every constructed segment with $r_{t+\tau}-\overline{r}$ where $\{r_{t}\}$ is a generated time series governed by a white noise process that is independent from $X$ and follows the same empirical distribution as $Y$ and $\overline{r}$ denotes the mean of $\{r_{t}\}$. We construct a collection of new segments $s_{t,\tau }^{D,\text{rand}}$
\begin{equation*}
s_{t,\tau }^{D,\text{rand}}=\left(x_{t-(D-2)}-\overline{x},\ldots,x_{t-1}-\overline{x},x_{t}-\overline{x},r_{t+\tau}-\overline{r}\right),  \quad t=D+\tau-1,D+\tau,\ldots,N-\tau.
\end{equation*} 
After that, by comparing the magnitudes of the entries in the constructed vectors $s_{t,\tau }^{D}$ and $s_{t,\tau }^{D,\text{rand}}$, we map every segment onto one of the $D!$ distinct ordinal patterns. We count the number of $s_{t,\tau }^{D}$ and $s_{t,\tau }^{D, \text{rand}}$ belonging to each ordinal pattern and divide them by the total number of constructed segments to compute the probabilities of ordinal patterns
\begin{equation*}
p_{\tau }\left( \pi _{i}\right) =\frac{\#\left\{s_{t,\tau }^{D}|s_{t,\tau
}^{D}\text{ has ordinal pattern }\pi _{i}\right\}}{N-D+2-\tau }%
,
\]\
\[
p^{\text{rand}}_{\tau }\left( \pi _{i}\right) =\frac{\#\left\{s_{t,\tau }^{D, \text{rand}}|s_{t,\tau
}^{D, \text{rand}}\text{ has ordinal pattern }\pi _{i}\right\}}{N-D+2-\tau }%
, 
\end{equation*}
where $(\pi_{i}; i=1,2,\ldots,D!)$ denotes the distinct ordinal patterns. An example may better illustrate the idea of an ordinal pattern. Let us take $D=3$. If the segment length is chosen to be 3, there are 6 possible ordinal patterns, which are given in Table~\ref{ordinal patterns} where $a$, $b$ and $c$ represent the first, second and last entry in the constructed segment.
\begin{table}[!htbp]
\centering
\begin{tabular}{@{}lll@{}}
\toprule
Ordinal Pattern &  & Condition \\
\midrule 
$\pi _{1}$ & $(3 2 1)$ & $a>b>c,$ \\ 
$\pi _{2}$ & $(3 1 2)$ & $a>c>b,$\\ 
$\pi _{3}$ & $(2 3 1)$ & $b>a>c,$\\ 
$\pi _{4}$ & $(2 1 3)$ & $c>a>b,$ \\ 
$\pi _{5}$ & $(1 3 2)$ & $b>c>a,$\\ 
$\pi _{6}$ & $(1 2 3)$ & $a<b<c.$\\
\bottomrule
\end{tabular}
\caption{Ordinal patterns for segment length $D=3$.}
\label{ordinal patterns}
\end{table}

\noindent The considered ordinal patterns do not include the cases where ties appear in the segments since in continuous time series dynamics, the probability of occurring equal values is negligible. Since the segments collection $(s_{t,\tau }^{D, \text{rand}}:t=D+\tau-1,\ldots,N-\tau)$ contains realizations from a simulated time series $\{r_{t}\}$, to reduce the uncertainty, we create the collection of segments $s_{t,\tau }^{D,\text{rand}}$ sufficient times (such as 500 times) to obtain an estimate of $\text{E}[p^{\text{rand}}_{\tau}(\pi_{i})]$ and $\text{s.d.}[p^{\text{rand}}_{\tau}(\pi_{i})]$. Measure $\text{K}(\bm{x}^{D-1}_{t},y_{t+\tau})$ is computed using
\begin{equation}\label{Kcomputation}
\text{K}(\bm{x}^{D-1}_{t},y_{t+\tau})=\sum_{\pi_{i}}\left(\dfrac{p_{\tau}(\pi_{i})-\text{E}[p^{\text{rand}}_{\tau}(\pi_{i})]}{\text{s.d.}[p^{\text{rand}}_{\tau}(\pi_{i})]}\right)^{2}.
\end{equation} 

By symbolizing the constructed segments according to their ordinal patterns, we convert the complex expression given in~\eqref{Ktheoretical} which involves an integral and comparison between an unknown continuous joint distribution and product of continuous marginal distributions into a simple comparison of easily accessible discrete distributions. The distribution of ordinal patterns on segments partitioned in the observed time series $(p_{\tau }\left( \pi _{i}\right); i=1,2,\ldots,D!)$ approximates the joint distribution between $\bm{x}^{D-1}_{t}$ and $y_{t+\tau}$. The expectation $\text{E}[p^{\text{rand}}_{\tau}(\pi_{i})]$ simulates the product of marginal distribution of $\bm{x}^{D-1}_{t}$ and $y_{t+\tau}$. The denominator $\text{s.d.}[p^{\text{rand}}_{\tau}(\pi_{i})]$ is included to standardize the contribution of each ordinal pattern. Additionally, since we only consider the relative magnitudes, the proposed measure is invariant to the marginal distribution of the variables under study, which is another desirable property of a dependence measure. Moreover, the ordinal-pattern-based measures also have the advantages of flexibly accounting for any form of structures, minimal requirement of prior assumptions and knowledge, and the robustness to outliers and both dynamical and stochastic noise \citep{bandt2002permutation}.

If $\bm{x}^{D-1}_{t}$ and $y_{t+\tau}$ are independent, we expect 
\begin{equation*}
p_{\tau}(\pi_{i})=\text{E}[p^{\text{rand}}_{\tau}(\pi_{i})], \qquad i=1,2,\ldots,D!.
\end{equation*}
In both cases the ordinal pattern in $\bm{x}^{D-1}_{t}$ has no impact, thus no predictive power for the upcoming $y_{t+\tau}$ or the randomly generated entry $r_{t+\tau}$ that is independent from $X$. Thereby the value of $\text{K}(\bm{x}^{D-1}_{t},y_{t+\tau})$ will be insignificant. On the other hand, if $\bm{x}^{D-1}_{t}$ affects the distribution of $y_{t+\tau}$, there will be a significant discrepancy between the level of regularities of ordinal patterns in the segments combining the the observations in $\bm{x}^{D-1}_{t}$ and $y_{t+\tau}$ and that involves an independent random variable. Thereby $\text{K}(\bm{x}^{D-1}_{t},y_{t+\tau})$ will be greater than zero. According to the specification given in~\eqref{Kcomputation}, the domain of the measure $\text{K}(\bm{x}^{D-1}_{t},y_{t+\tau})$ is $[0,+\infty)$. The greater its value the stronger dependence it indicates.

Given an observed time series $\{x_{t};t=1,\cdots,N\}$, to evaluate a given prediction model's point forecast performance, our statistics are
\begin{equation*}
\text{K}(\widehat{\bm{\varepsilon}}^{D-1}_{t},\widehat{\varepsilon}_{t+\tau}) \quad \text{and} \quad \text{K}(\bm{x}^{D-1}_{t},\widehat{\varepsilon}_{t+\tau})
\end{equation*}
where $\{\widehat{\varepsilon}_{t};t=1,\cdots,N\}$ represents the residuals formed by the difference between the observations and the constructed point forecasts from any model, $\widehat{\bm{\varepsilon}}^{D-1}_{t}=(\widehat{\varepsilon}_{t-D+2},\dots,\widehat{\varepsilon}_{t-1},\widehat{\varepsilon}_{t})$ contains $D-1$ historical entries of residuals and
$\bm{x}^{D-1}_{t}=(x_{t-D+2},\dots,x_{t-1},x_{t})$ collects the same number of historical prediction time series observations. Statistics 
$\text{K}(\widehat{\bm{\varepsilon}}^{D-1}_{t},\widehat{\varepsilon}_{t+\tau})$ and $\text{K}(\bm{x}^{D-1}_{t},\widehat{\varepsilon}_{t+\tau})$ quantify the extent to which the past residuals affect the current residual and past observations of the prediction target affect the current residual respectively. Their significance and relative order are informative in unveiling the sufficiency of the point forecast performance of the employed model. 

From the generalized form of time series dynamics in~\eqref{eqn:eqnTimeSeriesGForm}, the residuals generated from a prediction $\widehat{y_{t}}=\widehat{c}+\widehat{g}(x_{t-1},x_{t-2},...,x_{t-\tau_{p}})$ can be written in the form 
\begin{equation} \label{eqn:residual1}
\widehat{\varepsilon}_{t}=c^{r}+g^{r}(\bm{x}^{\tau^{*}}_{t-1})+\varepsilon_{t},
\end{equation}
where $\bm{x}^{\tau^{*}}_{t-1}=(x_{t-1},x_{t-2},...,x_{t-\tau^{*}})$, $\tau^{*}=\max(\tau_{T},\tau_{p})$, $\tau_{T}$ is the furthest lag in which past entries affects the current entry $x_{t}$, $\tau_{p}$ is the furthest lag used in the prediction, $c^{r}=c-\widehat{c}$ and $g^{r}(\cdot)=g(\cdot)-\widehat{g}(\cdot)$. If the predictor is equal to the oracle point predictor $\widehat{x}_{t}^{\text{oracle}}$, the resulting residuals are equal to the \enquote{true} additive innovations, i.e. 
\begin{equation} \label{eqn:residual2}
\widehat{\varepsilon}_{t}=\varepsilon_{t}.
\end{equation} 
Otherwise, the estimated residual contains three components, which are the remaining constant term, the residual deterministic function of past observations of $x_{t}$ and the \enquote{true} additive innovation term as specified in~\eqref{eqn:residual1}. 

Consequently, by substituting $\widehat{\varepsilon}_{t}$ with the generalized form of residuals in~\eqref{eqn:residual1} and~\eqref{eqn:residual2} and neglecting the constant term, under the condition that the point forecast is oracle,
\begin{align*}
\text{K}(\widehat{\bm{\varepsilon}}^{D-1}_{t},\widehat{\varepsilon}_{t+\tau})&\equiv
\text{Dependence}(\bm{\varepsilon}^{D-1}_{t},\varepsilon_{t+\tau})\\
\text{K}(\bm{x}^{D-1}_{t},\widehat{\varepsilon}_{t+\tau})&\equiv\text{Dependence}(\bm{x}^{D-1}_{t},\varepsilon_{t+\tau}).
\end{align*}
Otherwise
\begin{align*}
\text{K}(\widehat{\bm{\varepsilon}}^{D-1}_{t},\widehat{\varepsilon}_{t+\tau})&\equiv\text{Dependence}\{[g^{r}(\bm{x}^{\tau^{*}}_{t-D+1})+\varepsilon_{t-D+2},\ldots,g^{r}(\bm{x}^{\tau^{*}}_{t-1})+\varepsilon_{t}],g^{r}(\bm{x}^{\tau^{*}}_{t+\tau-1})+\varepsilon_{t+\tau}\},\\
\text{K}(\bm{x}^{D-1}_{t},\widehat{\varepsilon}_{t+\tau})&\equiv\text{Dependence}[\bm{x}^{D-1}_{t},g^{r}(\bm{x}^{\tau^{*}}_{t+\tau-1})+\varepsilon_{t+\tau}].
\end{align*}

Under the oracle point forecast there are two possible scenarios. When the innovation terms are independent of each other, both $\text{K}(\widehat{\bm{\varepsilon}}^{D-1}_{t},\widehat{\varepsilon}_{t+\tau})$ and $\text{K}(\bm{x}^{D-1}_{t},\widehat{\varepsilon}_{t+\tau})$ will be insignificant. This is due to the fact that no dependence structure exist between past and current innovations or between any historical observation in $\{x_{t}\}$ and the current innovation. However, when the innovation is dependent, both statistics $\text{K}(\widehat{\bm{\varepsilon}}^{D-1}_{t},\widehat{\varepsilon}_{t+\tau})$ and $\text{K}(\bm{x}^{D-1}_{t},\widehat{\varepsilon}_{t+\tau})$ will be significant. Their significance arises from the temporal dependence within $\{\varepsilon_{t}\}$. Besides, the value of $\text{K}(\bm{x}^{D-1}_{t},\widehat{\varepsilon}_{t+\tau})$ will be smaller than $\text{K}(\widehat{\bm{\varepsilon}}^{D-1}_{t},\widehat{\varepsilon}_{t+\tau})$ since the only connection between $\bm{x}^{D-1}_{t}$ and $\varepsilon_{t+\tau}$ is through the intermediate of the temporal dependence of $\{\varepsilon_{t}\}$ quantified by $\text{K}(\bm{\varepsilon}^{D-1}_{t},\varepsilon_{t+\tau})$. Assuming the temporal dependence within innovation process $\{\varepsilon_{t}\}$ deteriorates for increasing lags, the expected value and the relation between $\text{K}(\widehat{\bm{\varepsilon}}^{D-1}_{t},\widehat{\varepsilon}_{t+\tau})$ and $\text{K}(\bm{x}^{D-1}_{t},\widehat{\varepsilon}_{t+\tau})$ under different scenarios are summarized in Table~\ref{tab:KKYtest}. 

\begin{table}[htbp]
\tabcolsep 0.2in
\centering
\caption{Expected behaviours of statistic $\text{K}(\widehat{\bm{\varepsilon}}^{D-1}_{t},\widehat{\varepsilon}_{t+\tau})$ and $\text{K}(\bm{x}^{D-1}_{t},\widehat{\varepsilon}_{t+\tau})$ when the evaluated point forecast is oracle and inadequate, in the presence of independent and dependent innovations.}\label{tab:KKYtest} 
\begin{tabular}{@{}l>{\raggedright}p{5.6cm}p{7cm}@{}}
\toprule
     Point forecast                  &$\varepsilon_{t}$ is independent                                                                                      & $\varepsilon_{t}$ is dependent                                                                                                          \\ \midrule
Oracle &                                                                                    $\text{K}(\widehat{\bm{\varepsilon}}^{D-1}_{t},\widehat{\varepsilon}_{t+\tau})=0$ and $\text{K}(\bm{x}^{D-1}_{t},\widehat{\varepsilon}_{t+\tau})=0$ & $\text{K}(\widehat{\bm{\varepsilon}}^{D-1}_{t},\widehat{\varepsilon}_{t+\tau})>0$, $\text{K}(\bm{x}^{D-1}_{t},\widehat{\varepsilon}_{t+\tau})>0$ and $\text{K}(\widehat{\bm{\varepsilon}}^{D-1}_{t},\widehat{\varepsilon}_{t+\tau})$ $\geq$ $\text{K}(\bm{x}^{D-1}_{t},\widehat{\varepsilon}_{t+\tau})$                     \\ \hline
Inadequate & $\text{K}(\widehat{\bm{\varepsilon}}^{D-1}_{t},\widehat{\varepsilon}_{t+\tau})>0$, $\text{K}(\bm{x}^{D-1}_{t},\widehat{\varepsilon}_{t+\tau})>0$ and $\text{K}(\bm{x}^{D-1}_{t},\widehat{\varepsilon}_{t+\tau})>\text{K}(\widehat{\bm{\varepsilon}}^{D-1}_{t},\widehat{\varepsilon}_{t+\tau})$& $\text{K}(\widehat{\bm{\varepsilon}}^{D-1}_{t},\widehat{\varepsilon}_{t+\tau})>0$, $\text{K}(\bm{x}^{D-1}_{t},\widehat{\varepsilon}_{t+\tau})>0$ and $\text{K}(\widehat{\bm{\varepsilon}}^{D-1}_{t},\widehat{\varepsilon}_{t+\tau})$ can be less or greater than $\text{K}(\bm{x}^{D-1}_{t},\widehat{\varepsilon}_{t+\tau})$ \\ \bottomrule
\end{tabular}
\end{table}

Based on the relations specified above, we propose a point forecast sufficiency test named the PE model sufficiency test. Suppose the data generating process of an observed time series $\{x_{t};t=1,2,\ldots,N\}$ can be written in the form 
\begin{equation*}
x_{t}=c+g(x_{t-\tau},x_{t-\tau-1},\ldots)+\varepsilon_{t},
\end{equation*}
where $\text{E}(\varepsilon_{t})=0$, $\text{E}(\varepsilon_{t}|F_{t-\tau})=0$ and $F_{t-\tau}$ denotes the information set available at time $t-\tau$. Therefore $c+g(x_{t-\tau},x_{t-\tau-1},\ldots)$ constitutes the oracle $\tau$-step-ahead point forecast of $x_{t}$. A given model provides a $\tau$-step-ahead point forecast $\widehat{x}_{\tau,t}$ to predict $x_{t}$, and $\widehat{x}_{\tau,t}$ can be written in the form
\begin{equation*}
\widehat{x}_{\tau,t}=\widehat{c}+\widehat{g}(x_{t-\tau},x_{t-\tau-1},\ldots).
\end{equation*}
The null hypothesis of the PE model sufficiency test is
\begin{equation*}
\mathcal{H}_{0}: \widehat{g}(\cdot)=g(\cdot),
\end{equation*}
versus the alternative hypothesis that
\begin{equation*}
\mathcal{H}_{1}: \widehat{g}(\cdot)\neq g(\cdot).
\end{equation*}
The test assesses whether the deterministic relation postulated by the predictor coincides with the true deterministic function governing the first-order moment of the investigated process.   
The acceptance condition is
\begin{align}\label{eqn:KKYA}
\text{K}(\widehat{\bm{\varepsilon}}^{D-1}_{t},\widehat{\varepsilon}_{t+\tau})=0 \quad \text{and} \quad \text{K}(\bm{x}^{D-1}_{t},\widehat{\varepsilon}_{t+\tau})=0,
\end{align}   
and the rejection criteria of $H_{0}$ is
\begin{align}\label{eqn:KKYB}
\text{K}(\bm{x}^{D-1}_{t},\widehat{\varepsilon}_{t+\tau})>0 \quad\text{and} \quad \text{K}(\bm{x}^{D-1}_{t},\widehat{\varepsilon}_{t+\tau})>\text{K}(\widehat{\bm{\varepsilon}}^{D-1}_{t},\widehat{\varepsilon}_{t+\tau}).
\end{align}
The acceptance condition specified in~\eqref{eqn:KKYA} corresponds to the case summarized in Table~\ref{tab:KKYtest} when the point forecast is oracle and $\varepsilon_{t}$ are independent. The rejection criteria in~\eqref{eqn:KKYB} eliminate all possible scenarios of oracle point forecast summarized in Table~\ref{tab:KKYtest}. 

To determine whether $\text{K}(\widehat{\bm{\varepsilon}}^{D-1}_{t},\widehat{\varepsilon}_{t+\tau})$ and $\text{K}(\bm{x}^{D-1}_{t},\widehat{\varepsilon}_{t+\tau})$ are significantly greater than zero, we need to compare their values with the critical value of $\text{K}(\bm{x}^{D-1}_{t},y_{t+\tau})$ when time series $\{x_{t}\}$ and $\{y_{t}\}$ are independent. Since the considered measure is ordinal-based, it neglects and is invariant to the empirical distribution of the investigated series. The statistical property of $\text{K}(\bm{x}^{D-1}_{t},y_{t+\tau})$ is identical for any pair of independent time series with any form of continuous marginal distribution. Therefore, an estimate of the 95\%  
C.I. of $\text{K}(\bm{x}^{D-1}_{t},y_{t+\tau})$ can be obtained by taking the average of the 95th percentile of the measure computed on a sufficient number of paths of a pair of simulated independent and iid series. The generated series needs to be of the same length as the investigated time series and can be specified to follow any continuous distribution.
 
Apart from assessing the significance of $\text{K}(\widehat{\bm{\varepsilon}}^{D-1}_{t},\widehat{\varepsilon}_{t+\tau})$ and $\text{K}(\bm{x}^{D-1}_{t},\widehat{\varepsilon}_{t+\tau})$, the test needs to determine whether measure $\text{K}(\bm{x}^{D-1}_{t},\widehat{\varepsilon}_{t+\tau})$ is significantly greater than $\text{K}(\widehat{\bm{\varepsilon}}^{D-1}_{t},\widehat{\varepsilon}_{t+\tau})$ when $\text{K}(\bm{x}^{D-1}_{t},\widehat{\varepsilon}_{t+\tau})$ shows a significant value. However, when the independence assumptions cannot be met, the distribution of $\text{K}(\widehat{\bm{\varepsilon}}^{D-1}_{t},\widehat{\varepsilon}_{t+\tau})$ and $\text{K}(\bm{x}^{D-1}_{t},\widehat{\varepsilon}_{t+\tau})$, especially their variances vary depending on the data generating process of $\{\widehat{\varepsilon}_{t}\}$ and $\{x_{t}\}$. Therefore we choose to use a block bootstrapping method to estimate the critical values of the test statistics $\text{K}(\bm{x}^{D-1}_{t},\widehat{\varepsilon}_{t+\tau})-\text{K}(\widehat{\bm{\varepsilon}}^{D-1}_{t},\widehat{\varepsilon}_{t+\tau})$ to conclude its significance when $\text{K}(\bm{x}^{D-1}_{t},\widehat{\varepsilon}_{t+\tau})$ exceeds its critical value under independence assumptions. 

The block bootstrap requires the selection of block length. We choose $l=20$ to be the block length used in the subsequent simulation studies and empirical analysis. The selected block length $l=20$ is approximately equal to $N^{1/3}$, where $N$ is the sample size. There is a trade-off in selecting the block length. The length of the block should be long enough to preserve the temporal dependence structure originally present in the observed data, in the meantime, ensure the sufficient number of blocks partitioned from the time series. Given that $\tau=20$ is the maximum lag that significant autocorrelation structures present in our investigated time series (as indicated in Figure~\ref{fig4}), we choose 20 to be the bootstrap block length to estimate the distribution of test statistics in our proposed test. We also carried out a sensitivity analysis to examine the impact of the block length on our proposed test results. We simulated 500 paths of GARCH innovations from the same specification as $X3$, and computed the distribution of $K(\widehat{\bm{\epsilon}}_t^{D-1}, \widehat{\epsilon}_{t+\tau})$ on the simulated innovations with the estimated distribution of $K(\widehat{\bm{\epsilon}}_t^{D-1}, \widehat{\epsilon}_{t+\tau})$ from the block bootstrap on one realization of the innovation. They are expected in close resemblance. We repeated the same procedures using the block length of $N^{1/2}, N^{1/3}, N^{1/4}$ and $N^{1/5}$. The sensitivity analysis suggests that block bootstrap using block length of $N^{1/2}, N^{1/3}$ performs much better than using block length of $N^{1/4}$ and $N^{1/5}$. The result stresses choosing the block length that preserves the temporal dependence structure present in the observed data. Further, when the block length is within a reasonable range, the performance of the proposed test is insensitive to various selections of the block length.

One might notice, the acceptance and rejection conditions given in~\eqref{eqn:KKYA} and~\eqref{eqn:KKYB} do not exhaustively cover all possible scenarios. There is one particular scenario when we cannot ascertain whether the constructed predictor is sufficient or not. That is when 
\begin{align}\label{eqn:inconclusive}
\text{K}(\bm{x}^{D-1}_{t},\widehat{\varepsilon}_{t+\tau})>0 \quad \text{and} \quad \text{K}(\bm{x}^{D-1}_{t},\widehat{\varepsilon}_{t+\tau})<\text{K}(\widehat{\bm{\varepsilon}}^{D-1}_{t},\widehat{\varepsilon}_{t+\tau}). 
\end{align}
Two possible cases can lead to the above outcome. First, the point forecast is sufficient, but the innovation exhibits strong dependence structures. Second, the prediction is inadequate. However, the dependence structure within innovations is too strong that overweights the remaining deterministic relations' in the original dynamics. Therefore when the test statistics is as in~\eqref{eqn:inconclusive}, our test cannot make an affirmative conclusion of the constructed predictor's sufficiency. This limitation is a weakness of our proposed test where future studies can be carried out.  

Our proposed PE model sufficiency test requires no prior information or assumptions about the system underlying the observed time series or the model that makes predictions, but only two pre-chosen parameters: the segment length $D$ and the delay $\tau$. Segment length $D$ determines the number of entries contained in every constructed segment, thereby the total number of possible ordinal patterns. Clearly, with more ordinal patterns, the measure is more capable of capturing the complex dynamics underlying the observed data. The choice of delay $\tau $ provides the flexibility to investigate the structure of time series over short-term and long-term dynamics, just like the lag in the autocorrelation. 

As for the common choice of $D$, we follow the guidelines of choosing the segment length in the PE where the test statistic $\text{K}$ inherits the ordinal concept from. According to \cite{bandt2002permutation}, $D$ is recommended to be chosen in the range of $3\leq D\leq 7$. The choice of parameter $D$ needs to be subject to the restriction: $N-\left(D-1\right) \tau \gg D!$ \citep{rosso2007distinguishing,kowalski2007bandt,zanin2008forbidden}. The rationale behind the recommended range of $D$ is that, since we partition the observed time series into numbers of overlapping segments and map the constructed segments into $D!$ possible ordinal patterns, the constructed number length, $N-\tau-1$, should excessively exceed the number of possible pattern categories. If data are relatively short with a sample size of 1000, we recommend $D=3$. If data are relatively long with a sample size of 10000, we recommend $D=4$ or 5. As for the choice of delay, for most financial dynamics, the temporal dependence structure is generally strongest between entries that are closest to each other. Therefore in the latter applications of our proposed test, we mainly evaluate the test statistics at $\tau=1$ to draw inferences about the sufficiency of the employed models.  

\section{Simulation studies} \label{sec:3}

We conduct simulation studies to demonstrate several applications of our newly proposed test. The PE model sufficiency test will be applied to six different simulated time series that correspond to different time series dynamics, all transformed into the general additive form of time series process given in~\eqref{eqn:eqnTimeSeriesGForm}. An overview of the six forms of simulating time-series dynamics is given below.  
\begin{itemize}
\item $X1$: linear deterministic function $g(\cdot)$ with iid normally distributed innovations $\varepsilon_{t}$ (ARMA)
\item $X2$: linear deterministic function $g(\cdot)$ with iid asymmetrically distributed innovations $\varepsilon_{t}$ 
\item $X3$: linear deterministic function $g(\cdot)$ with asymmetrically distributed structural innovations $\varepsilon_{t}$ (GARCH)
\item $X4$: nonlinear deterministic function $g(\cdot)$ with iid normally distributed innovations $\varepsilon_{t}$ 
\item $X5$: nonlinear deterministic function $g(\cdot)$ with iid asymmetrically distributed innovations $\varepsilon_{t}$
\item $X6$: nonlinear deterministic function $g(\cdot)$ with asymmetrically distributed structural innovations $\varepsilon_{t}$.
\end{itemize}

In the simulation studies, the PE model sufficiency test is used to indicate the sufficiency of various point forecasts constructed by different models. The prediction models considered include the ARMA model, GARCH model, Gaussian Process (GP) regression, and support vector regression (SVR). The specifications and the parameter estimation procedures of the employed models can be found in \cite{ahmed2010empirical} and \cite{tsay2005analysis}. Each simulated series is of length 6360, which corresponds to the number of one-hour intervals of a one-year financial time series excluding weekends. The selected data length mimics the size of a one-year-long one-hour interval empirical financial volatility series that we analyze in the subsequent section. The first 5160 observations (around 80\% of total length) are used to estimate the model parameters, and the last 1200 observations are used in assessing out-of-sample prediction accuracy. The non-parametric approaches, namely GP and SVR, require the pre-chosen hyper-parameters. For non-parametric models, the first 4000 observations are used to train the prediction model. The subsequent 1160 observations are cross-validation set, which is used to select the optimal hyper-parameters and determine the number of lagged observations $d$ used as input variables. The hyper-parameters tuning and the selection of lag $d$ are through a heuristic grid search that minimizes the mean squared error (MSE) in the cross-validation set. The last 1200 observations are used in assessing out-of-sample prediction accuracy. 

The validity and credibility of our proposed test are verified by contrasting the inferences drawn from the PE model sufficiency test with the prediction error rate of each considered model. The prediction error rate reflects the true point forecast performance, and is measured by the distance between the constructed predictor $\widehat{x}_{t}$ from the oracle predictor $\widehat{x}^{\text{oracle}}_{t}$ of $x_{t}$ relative to $\widehat{x}^{\text{oracle}}_{t}$'s variance, i.e.
\begin{align*}
\text{prediction error rate} &=\frac{\frac{1}{N}\sum^{N}_{t=1}\left[\left(\widehat{x}_{t}-\widehat{x}^{\text{oracle}}_{t}\right)-\frac{1}{N}\sum^{N}_{t=1}\left(\widehat{x}_{t}-\widehat{x}^{\text{oracle}}_{t}\right)\right]^{2}}{\text{var}\left(\widehat{x}^{\text{oracle}}_{t}\right)}.
\end{align*}
We subtract the systematic bias $\frac{1}{N}\sum^{N}_{t=1}(\widehat{x}_{t}-\widehat{x}^{\text{oracle}}_{t})$ in evaluating point forecast performance because our ordinal-based sufficiency test cannot reveal the derivation in the estimation of the constant term. In other words, the systematic bias introduced by the predictor cannot be captured by the dependence measures used in our test.
Due to the nature of the simulation study, the data generating process of the simulated series is known in advance, unlike in empirical analysis. Therefore we can acquire the oracle point predictor for every simulated series. Based on the general form of time series given in~\eqref{eqn:eqnTimeSeriesGForm}
\begin{equation*}
\widehat{x}^{\text{oracle}}_{t}=\text{E}(x_{t}|F_{t-1})=x_{t}-\varepsilon_{t},
\end{equation*}
where $F_{t-1}$ denotes the information set available at time $t-1$, $\text{E}(\varepsilon_{t})=0$ and $\text{E}(\varepsilon_{t}|F_{t-1})=0$. $\widehat{x}^{\text{oracle}}_{t}$ is the best one-step-ahead forecast one can construct of $x_{t}$. 

Due to the inevitable uncertainties in the model estimation procedures, even the most optimal model would not entirely eliminate prediction error. We divide the prediction error rate into three categories to indicate the cases, namely \begin{inparaenum}
\item[1)] when the postulated point predictions are reasonably close; \item[2)] within a moderate distance; and \item[3)] significantly away from the oracle predictor relative to the overall variations of the underlying dynamics. \end{inparaenum} Based on the value of the prediction error rate, we classify the point prediction performance of the considered models into three categories: sufficient ($\text{prediction \ error \ rate}\leq5\%$), acceptable ($5\%<\text{prediction \ error \ rate}\leq 15\%$) and inadequate ($\text{prediction \ error \ rate}>15\%$).

In addition to examining our new test's validity, the objective of the simulation study is to investigate the point forecast capability of the considered models in response to various potentially challenging properties commonly observed in financial time series, such as non-normality, non-linearity, and dynamical innovations. The specifications of the data generating process of the six simulated time series are given below, all transformed into the additive form given in~\eqref{eqn:eqnTimeSeriesGForm}
\begin{itemize}
\item $X1$: simulated ARMA(1,1) series with parameter $(\phi_{0},\phi_{1},\theta_{1},\delta^{2})=(0.18,0.9,0.74,9)$ . The deterministic function of the data generating process is $g(x_{t-1},x_{t-2},...,x_{t-\tau_{T}})=0.69+\sum_{i=1}^{\infty}0.16\times 0.74^{i-1}x_{t-i}$. Innovation terms $\varepsilon_{t}$ are iid with normal distribution with mean 0 and variance 9;
\item $X2$: simulated ARMA(1,1) series with the same $g(\cdot)$ as in $X1$. However, the innovation terms $\varepsilon_{t}$ are not following a normal distribution. Instead, $\varepsilon_{t}$ follows the empirical asymmetric distribution as the unconditional distribution of the innovation of the simulated GARCH(1,1) series $X3$;
\item $X3$: simulated GARCH(1,1) series where the squared returns is the target objective, with parameter $(\alpha_{0},\alpha_{1},\beta_{1})=(0.18,0.16,0.74)$. The parameters of the GARCH model are selected so that the deterministic function $g(\cdot)$ in $X3$ is identical to that in $X1$ and $X2$. The innovation terms $\varepsilon_{t}$ are asymmetrically distributed and dependent on each other;
\item $X4$: a kernel function formed nonlinear $g(\cdot)$ with the same innovation term as in $X1$;
\item $X5$: same nonlinear $g(\cdot)$ as in $X4$ with the same innovation term as in $X2$;
\item $X6$: same nonlinear $g(\cdot)$ as in $X4$ with the innovation term as in $X3$.
\end{itemize}

\begin{figure}[!htbp]
\centering
\subfloat[]{
\includegraphics[width=8.2cm]{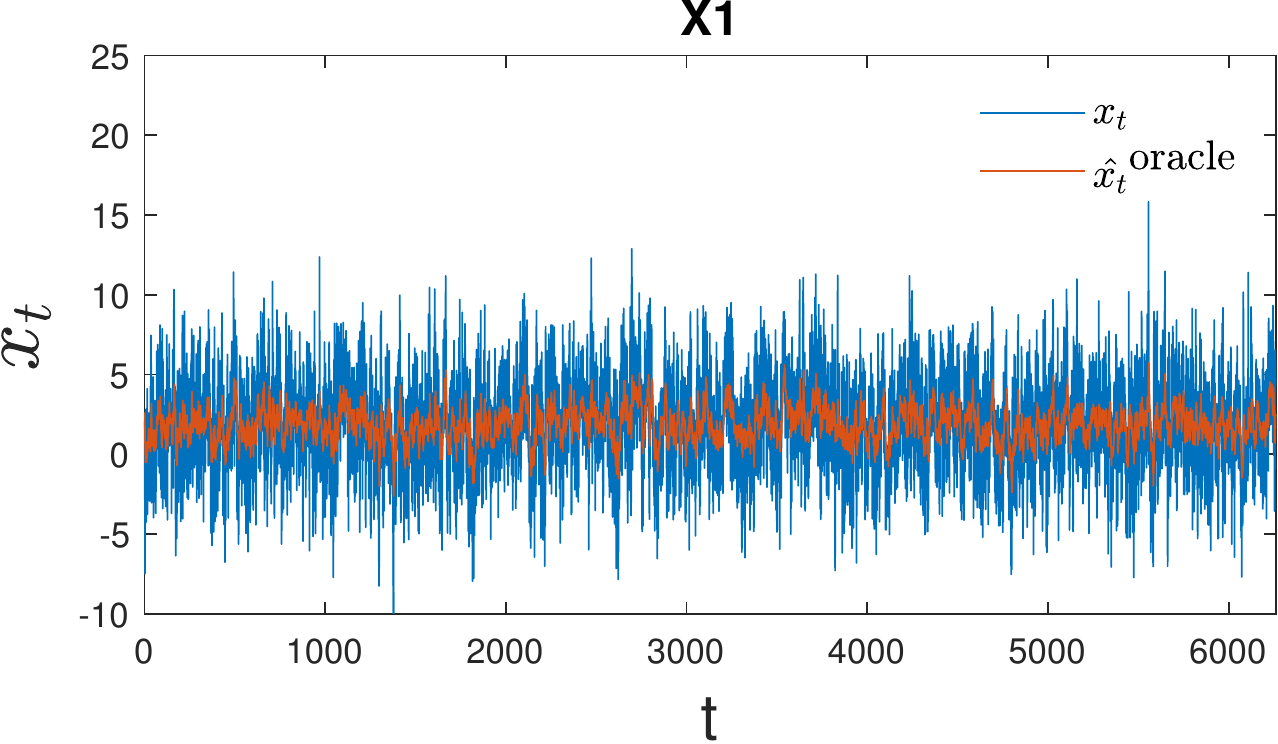}
\label{fig1:subfig1}}
\hfil
\subfloat[]{
\includegraphics[width=8.2cm]{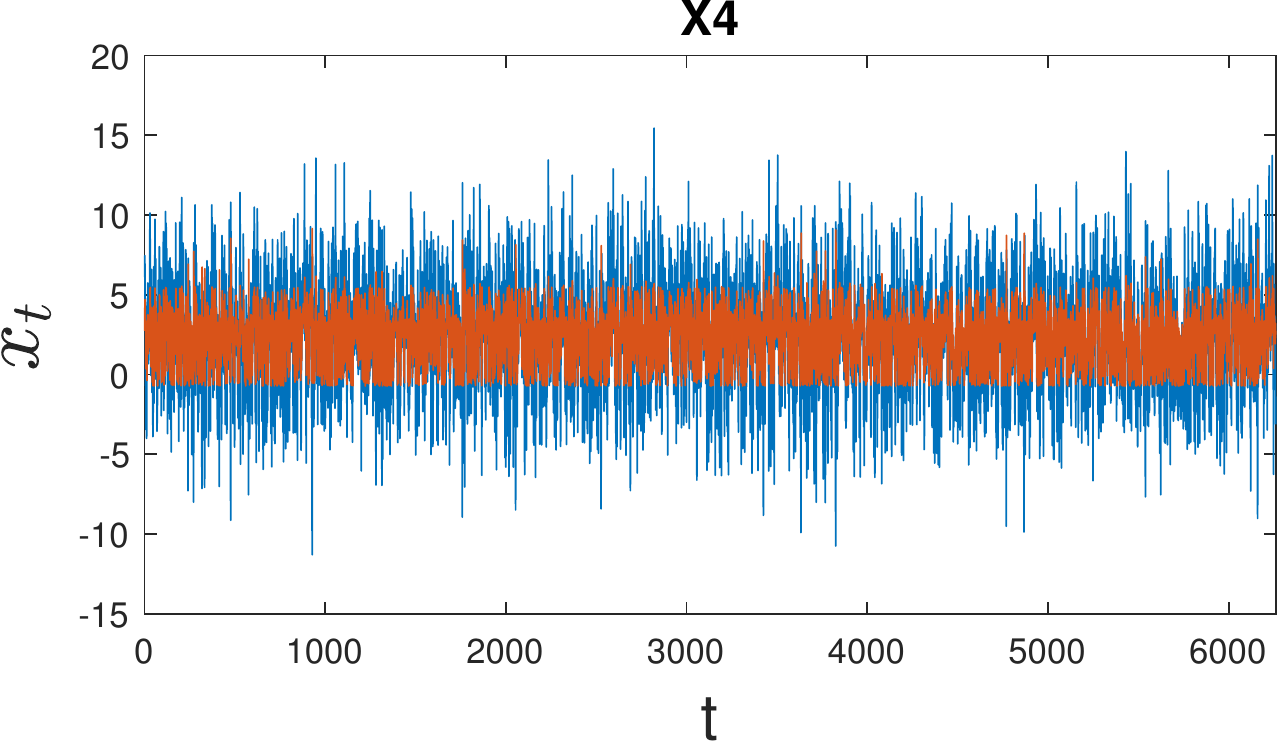}
\label{fig1:subfig4}}

\subfloat[]{
\includegraphics[width=8.2cm]{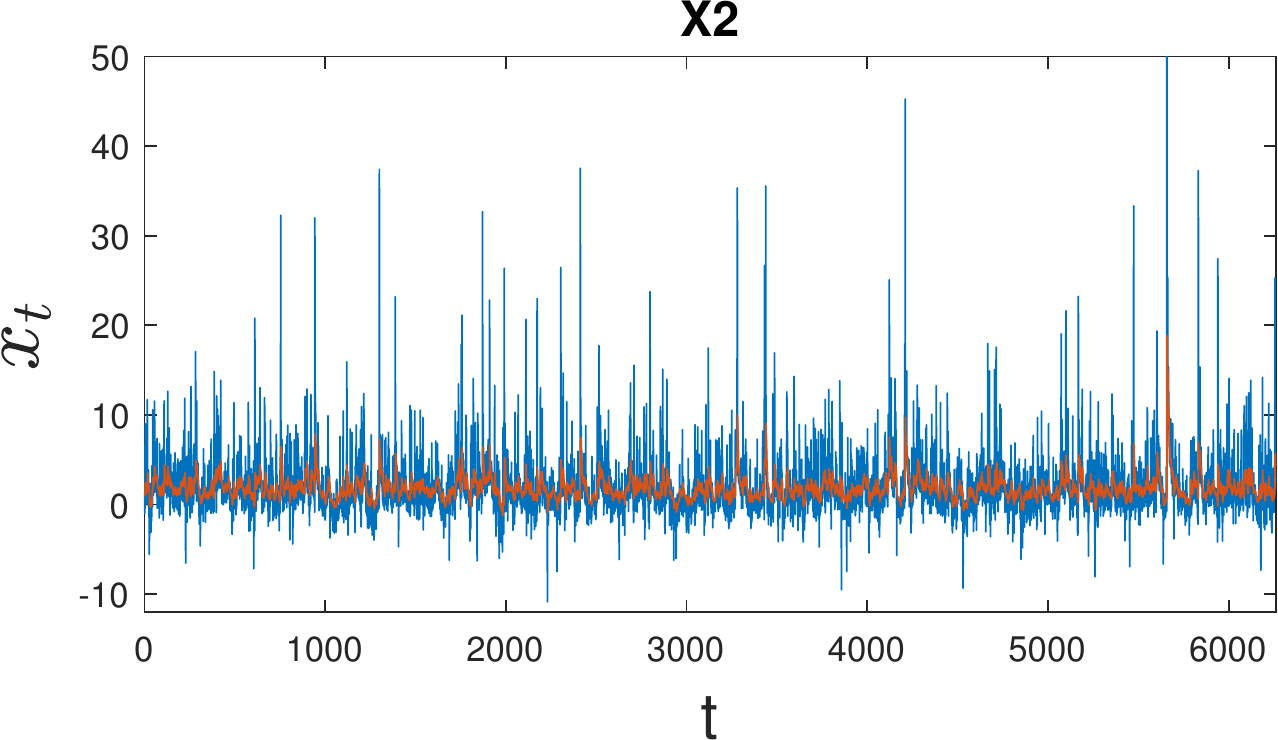}
\label{fig1:subfig2}}
\hfil
\subfloat[]{
\includegraphics[width=8.2cm]{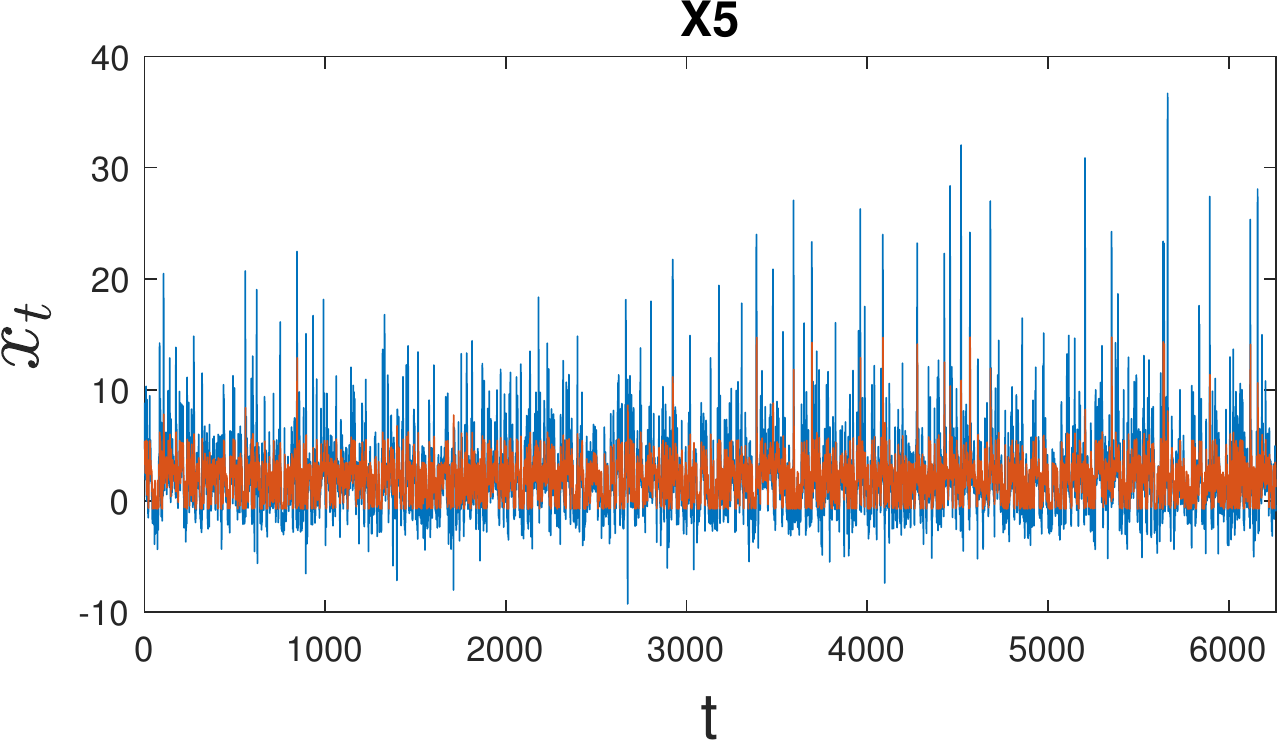}
\label{fig1:subfig5}}

\subfloat[]{
\includegraphics[width=8.2cm]{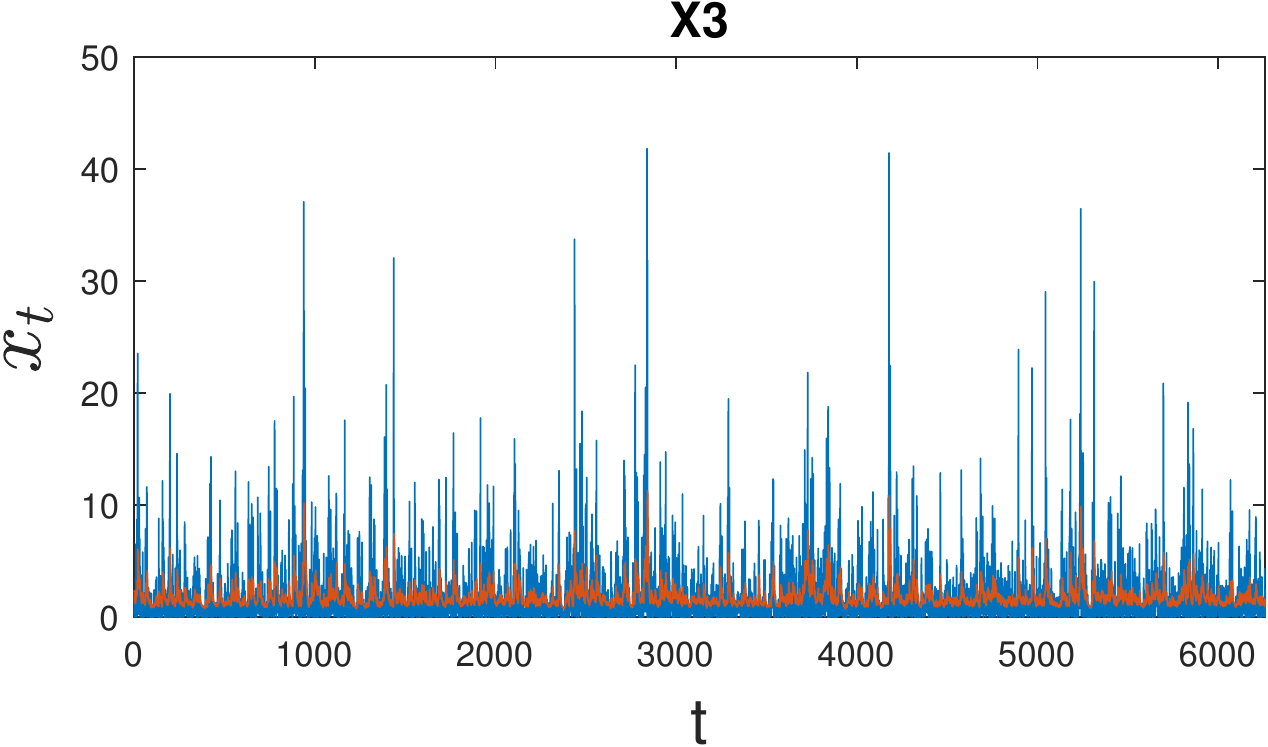}
\label{fig1:subfig3}}
\hfil
\subfloat[]{
\includegraphics[width=8.2cm]{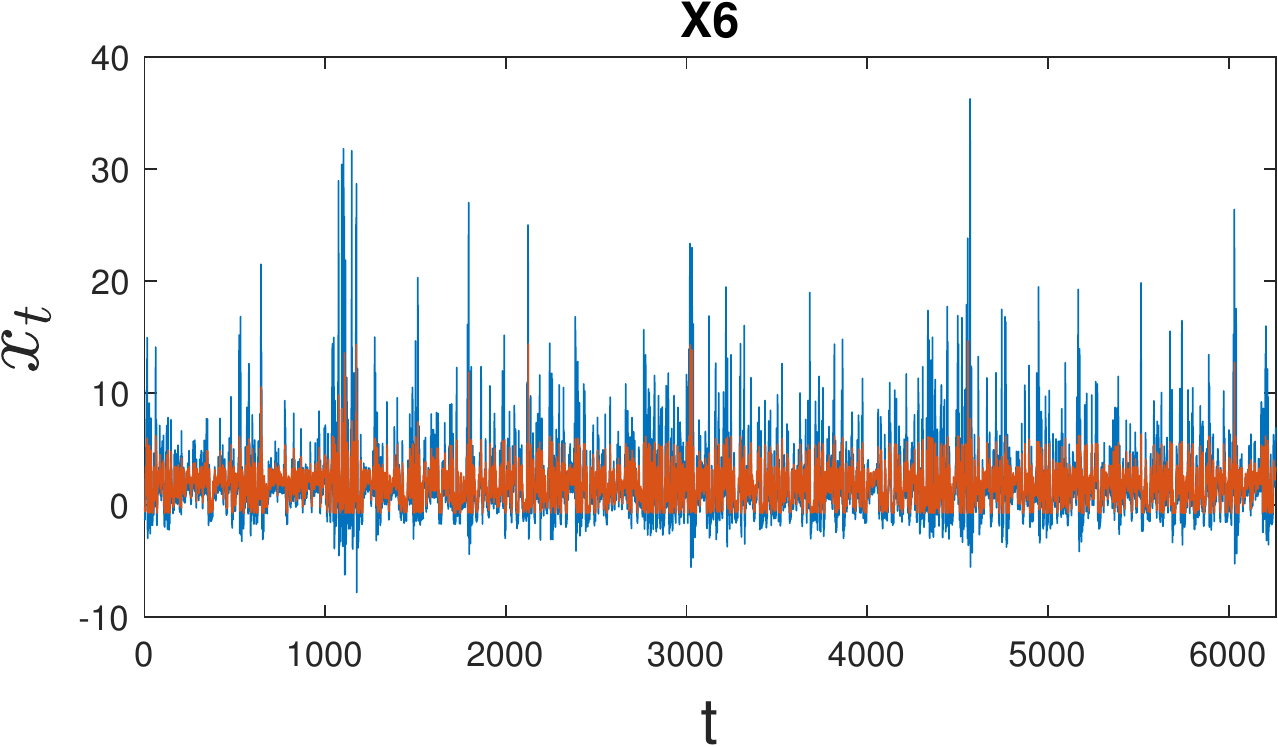}
\label{fig1:subfig6}}
\caption{Plot of simulated series $X1$ to $X6$ and their respective oracle one-step-ahead predictions.}
\label{fig1}
\end{figure}

\begin{figure}[htbp]
\centering
\subfloat[]{
\includegraphics[width=8cm]{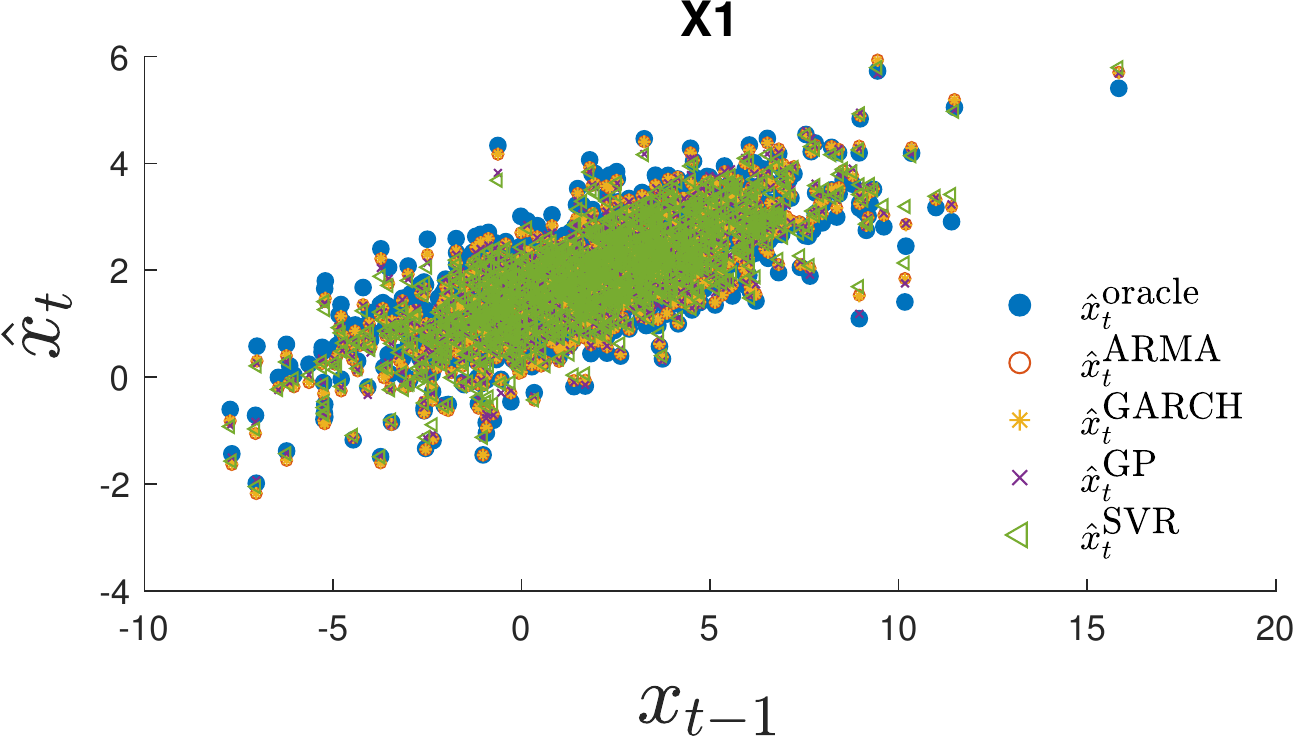}
\label{fig2:subfig1}}
\hfil
\subfloat[]{
\includegraphics[width=8cm]{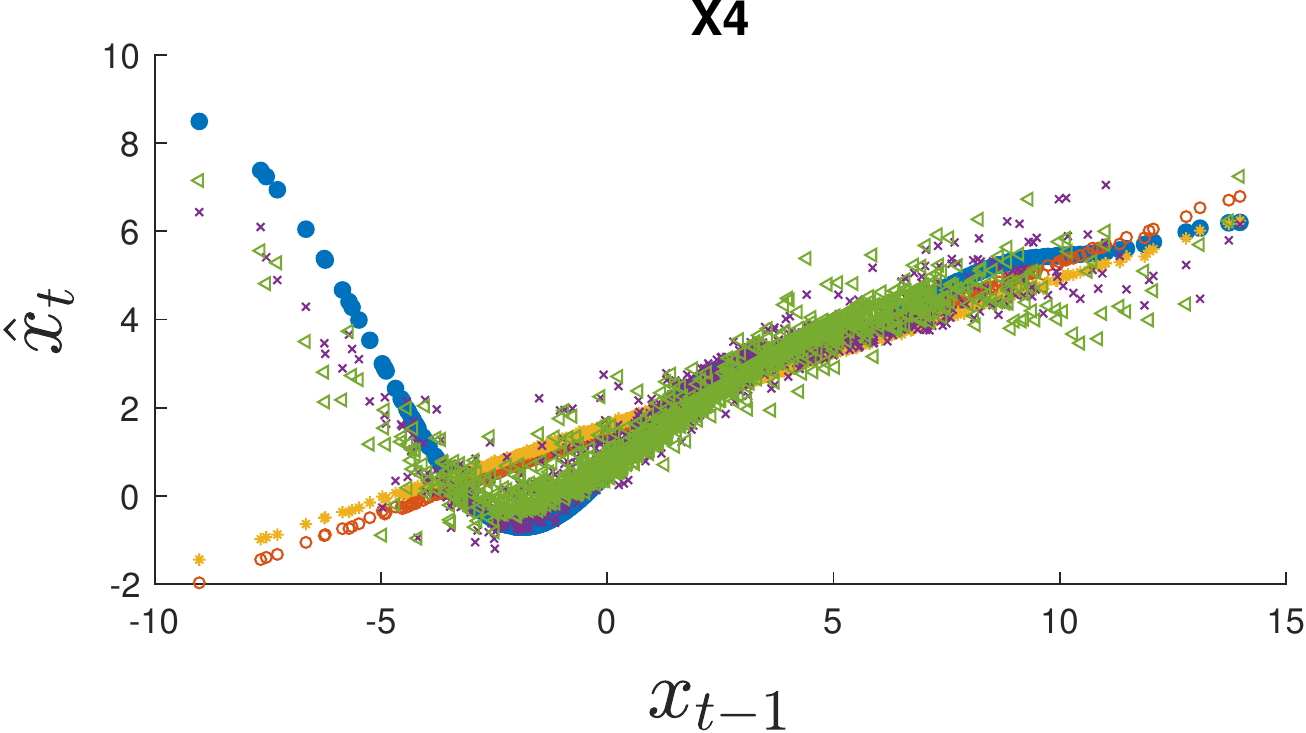}
\label{fig2:subfig2}}

\subfloat[]{
\includegraphics[width=8cm]{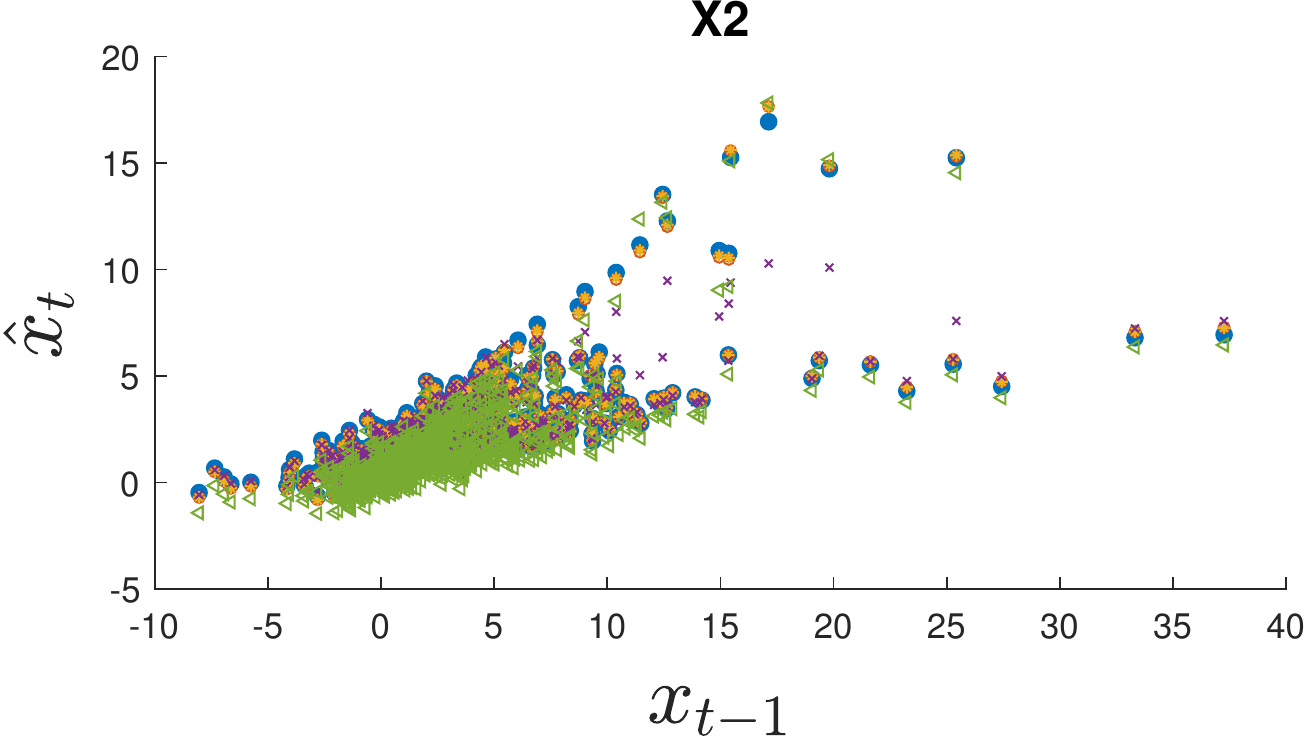}
\label{fig2:subfig3}}
\hfil
\subfloat[]{
\includegraphics[width=8cm]{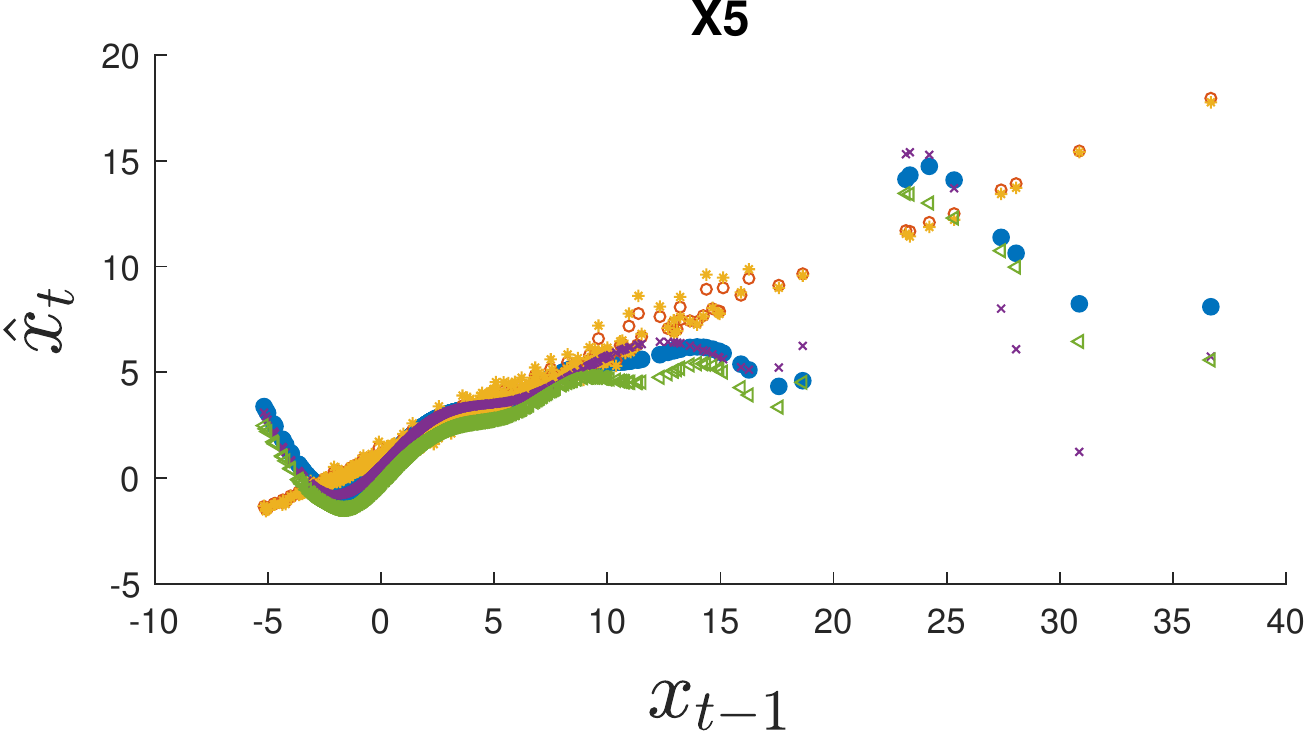}
\label{fig2:subfig4}}

\subfloat[]{
\includegraphics[width=8cm]{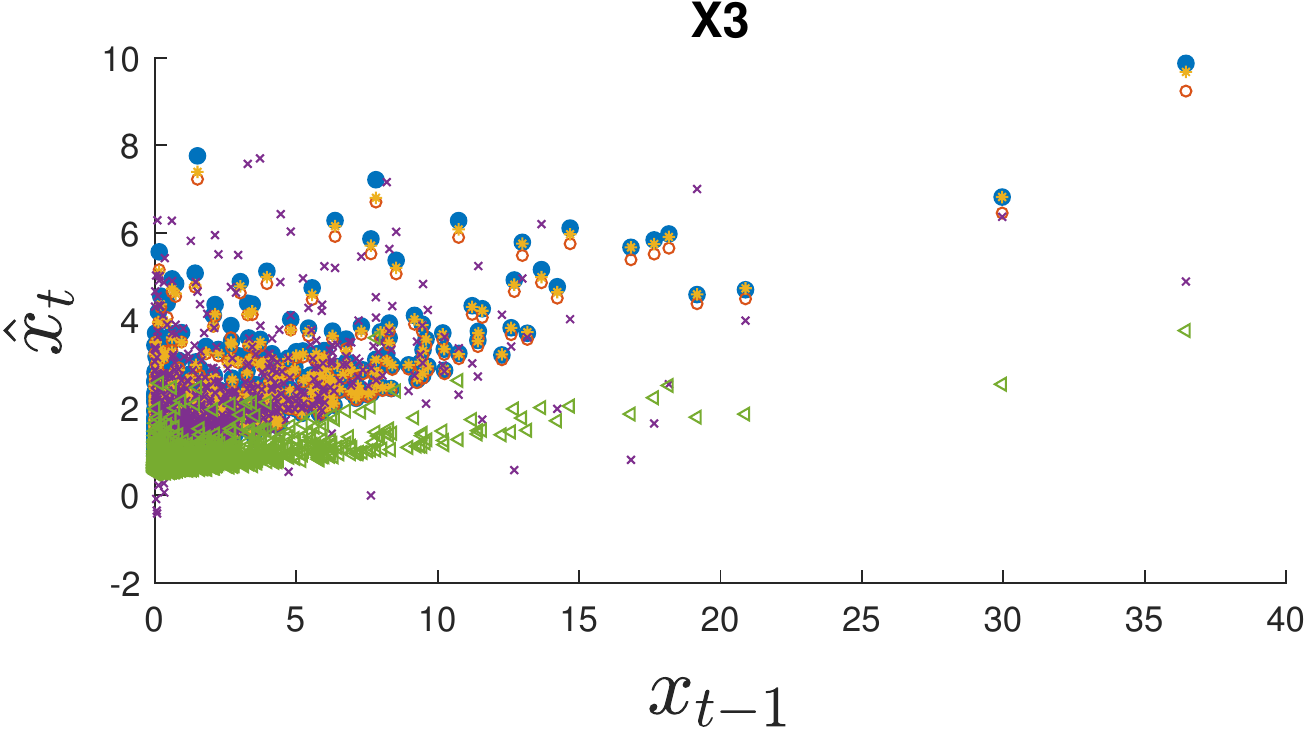}
\label{fig2:subfig5}}
\hfil
\subfloat[]{
\includegraphics[width=8cm]{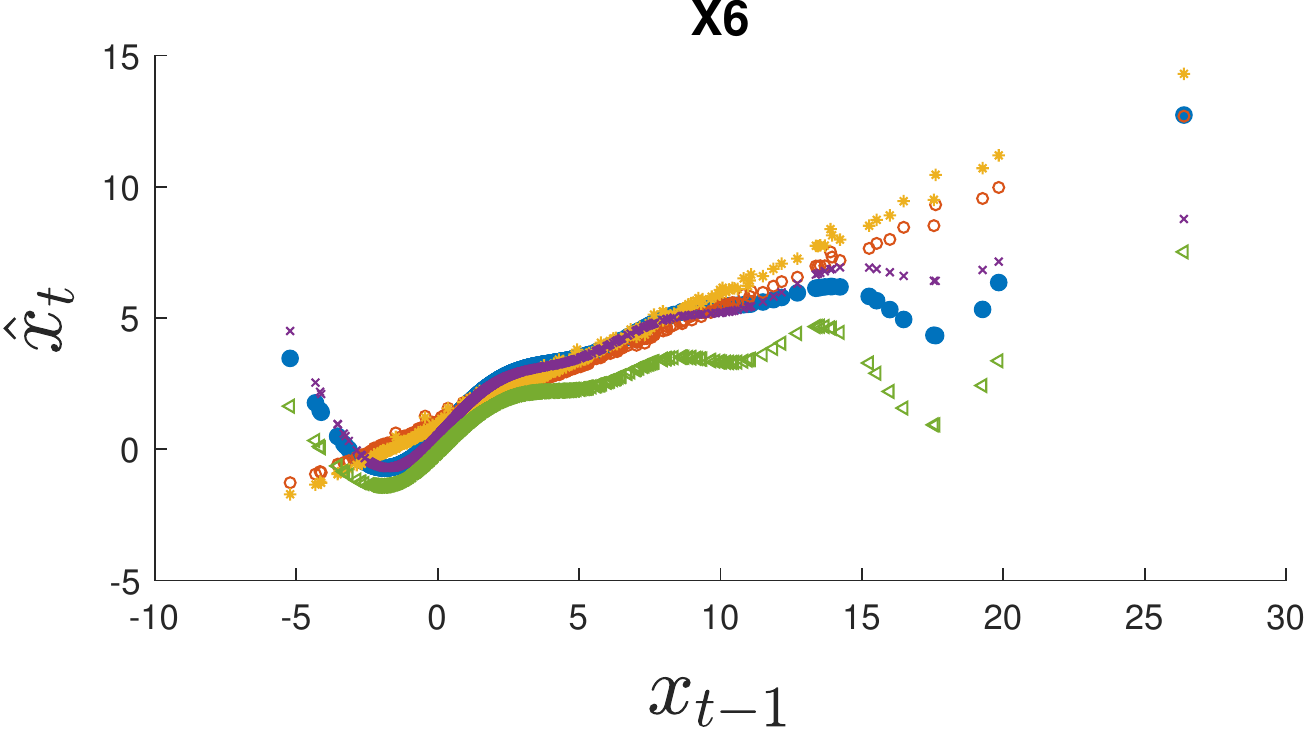}
\label{fig2:subfig6}}
\caption{Scatter plot of the oracle one-step-ahead predictions $\widehat{x}_{t}^{\text{oracle}}$ and prediction constructed by ARMA, GARCH, GP and SVR model fitted to each simulated series $X1$ to $X6$ against the nearest lagged observations $x_{t-1}$.}
\label{fig2}
\end{figure}

The six simulated time series specifications are motivated by the estimated models fitted to empirical volatility time series studied in the next section. By doing so, we expect the simulated time series to have similar characteristics as the real-world financial time series. Figure~\ref{fig1} plots the simulated series $X1$ to $X6$ and their respective oracle one-step-ahead point forecast $\widehat{x}^{\text{oracle}}_{t}$. Figure~\ref{fig2} provides the scatter plot to display the relation between the oracle point forecast $\widehat{x}^{\text{oracle}}_{t}$ and the nearest lagged observation $x_{t-1}$. Additionally, the distances between the prediction $\widehat{x}_{t}$ made from the considered models and the oracle point forecast are also revealed.  

All relevant statistics reflecting the point forecast performance of the considered models are summarized in Table~\ref{tb:KKYtestsimulation}. The first row records the signal-to-noise ratio of the simulated series to reflect their noise contamination level, i.e.
\begin{align*}
\text{signal-to-noise ratio}=\frac{\sum^{N}_{t=1}(\widehat{x}^{\text{oracle}}_{t})^2}{\sum^{N}_{t=1}(\varepsilon_{t}^{2})}.
\end{align*} 
Subsequently, the value of the prediction error rate is given to reflect the \enquote{true} level of the point forecast performance of the competing models. 
\begin{footnotesize}
\begin{longtable}{@{}lllllll@{}}
\caption{{\footnotesize Prediction performance ARMA, GARCH, GP and SVR models on six simulated time series generated through linear and nonlinear deterministic relation and with different form of innovations. Prediction error rate indicates the \enquote{true} predictor accuracy: $\text{prediction error rate}=\frac{1}{N}\sum^{N}_{t=1}\left[(\widehat{x}_{t}-\widehat{x}^{\text{oracle}}_{t})-\frac{1}{N}\sum^{N}_{t=1}(\widehat{x}_{t}-\widehat{x}^{\text{oracle}}_{t})\right]^{2}/\text{var}(\widehat{x}^{\text{oracle}}_{t})$. Based on the value of prediction error rate, the performance of various models are classified into three categories: sufficient ($\text{prediction \ error \ rate}\leq5\%$) cell filled by blue, acceptable ($5\%<\text{prediction \ error \ rate}\leq 15\%$) filled by yellow and inadequate ($\text{prediction \ error \ rate}>15\%$) filled by red. adjMSEratio measures the prediction accuracy without the knowledge of generating process: $\text{adjMSEratio}=\frac{1}{N}\sum^{N}_{t=1}\left[(x_{t}-\widehat{x}_{t})-\frac{1}{N}\sum^{N}_{t=1}(x_{t}-\widehat{x}_{t})\right]^{2}/\text{var}(x_{t})$. $\text{bias ratio}$ indicates the bias of each model introduces to the prediction relative to the variance of the investigated data: $\text{bias ratio}=\left[\frac{1}{N}\sum^{N}_{t=1}(x_{t}-\widehat{x}_{t})\right]^2/\text{var}(x_{t})$. $d$ denotes the optimized number of lagged observations used as input variables in constructing GP and SVR. $\text{K}(\widehat{\bm{\varepsilon}}^{D-1}_{t},\widehat{\varepsilon}_{t+\tau})$ and $\text{K}(\bm{x}^{D-1}_{t},\widehat{\varepsilon}_{t+\tau})$ with $D=4$ and $\tau=1$ are the statistics required in the PE model sufficient test, they are computed in both training set and test set. The inferences drawn from BDS test are also included to compared with that form the PE model sufficiency test.}}
\\\toprule
 &
  $X1$ &
  $X2$ &
  $X3$ &
  $X4$ &
  $X5$ &
  $X6$ \\ \midrule
\endfirsthead
 &
  $X1$ &
  $X2$ &
  $X3$ &
  $X4$ &
  $X5$ &
  $X6$ \\ \midrule
\endhead
\hline\multicolumn{7}{r}%
{{Continued on next page}} \\
\endfoot
\endlastfoot
signal-to-noise ratio &
  55.12\% &
  43.91\% &
  45.02\% &
  97.00\% &
  89.16\% &
  79.17\% \\ \midrule
\multicolumn{1}{c}{Oracle} &
   &
   &
   &
   &
   &
   \\ \midrule
prediction error rate &
  \cellcolor[HTML]{DAE8FC}0.00\% &
  \cellcolor[HTML]{DAE8FC}0.00\% &
  \cellcolor[HTML]{DAE8FC}0.00\% &
  \cellcolor[HTML]{DAE8FC}0.00\% &
  \cellcolor[HTML]{DAE8FC}0.00\% &
  \cellcolor[HTML]{DAE8FC}0.00\% \\
adjMSEratio &
  92.86\% &
  89.28\% &
  95.58\% &
  74.23\% &
  74.41\% &
  73.24\% \\
bias ratio &
  0.00\% &
  0.00\% &
  0.00\% &
  0.00\% &
  0.00\% &
  0.00\%  \\
$\text{K}(\widehat{\bm{\varepsilon}}^{D-1}_{t},\widehat{\varepsilon}_{t+\tau})$ in-sample &
  22.77 &
  31.60 &
  342.16 &
  28.83 &
  26.85 &
  307.90 \\
$\text{K}(\bm{x}^{D-1}_{t},\widehat{\varepsilon}_{t+\tau})$ in-sample &
  21.80 &
  28.89 &
  179.55 &
  29.44 &
  25.79 &
  89.98 \\
PE sufficient test &
  \cellcolor[HTML]{FFFFFF}{\color[HTML]{3166FF} Accept} &
  \cellcolor[HTML]{FFFFFF}{\color[HTML]{3166FF} Accept} &
  Inconclusive &
  {\color[HTML]{3166FF} Accept} &
  {\color[HTML]{3166FF} Accept} &
  Inconclusive \\
$\text{K}(\widehat{\bm{\varepsilon}}^{D-1}_{t},\widehat{\varepsilon}_{t+\tau})$ out-of-sample &
  30.31 &
  29.97 &
  78.00 &
  29.55 &
  21.34 &
  90.40 \\
$\text{K}(\bm{x}^{D-1}_{t},\widehat{\varepsilon}_{t+\tau})$ out-of-sample &
  30.64 &
  18.96 &
  36.11 &
  19.67 &
  22.39 &
  45.85 \\
PE sufficiency test &
  {\color[HTML]{3166FF} Accept} &
  {\color[HTML]{3166FF} Accept} &
  {\color[HTML]{3166FF} Accept} &
  {\color[HTML]{3166FF} Accept} &
  {\color[HTML]{3166FF} Accept} &
  Inconclusive \\
BDS test &
  Accept &
  Accept &
  Reject &
  Accept &
  Accept &
  Reject \\ \midrule
\multicolumn{1}{c}{ARMA} &
   &
   &
   &
   &
   &
   \\ \midrule
prediction error rate&
  \cellcolor[HTML]{DAE8FC}1.47\% &
  \cellcolor[HTML]{DAE8FC}0.52\% &
  \cellcolor[HTML]{DAE8FC}0.80\% &
  \cellcolor[HTML]{FE996B}32.17\% &
  \cellcolor[HTML]{FE996B}17.90\% &
  \cellcolor[HTML]{FE996B}16.48\% \\
adjMSEratio &
  93.00\% &
  89.26\% &
  95.15\% &
  83.12\% &
  78.45\% &
  79.75\% \\
bias ratio &
  0.00\% &
  0.12\% &
  0.00\% &
  0.00\% &
  0.04\% &
  0.00\%  \\
$\text{K}(\widehat{\bm{\varepsilon}}^{D-1}_{t},\widehat{\varepsilon}_{t+\tau})$ in-sample &
  22.71 &
  27.31 &
  305.43 &
  48.22 &
  135.00 &
  404.59 \\
$\text{K}(\bm{x}^{D-1}_{t},\widehat{\varepsilon}_{t+\tau})$ in-sample &
  25.61 &
  26.81 &
  174.23 &
  80.81 &
  155.13 &
  288.11 \\
PE sufficiency test &
  {\color[HTML]{3166FF} Accept} &
  {\color[HTML]{3166FF} Accept} &
  Inconclusive &
  {\color[HTML]{FE0000} Reject} &
  Inconclusive &
  Inconclusive \\
$\text{K}(\widehat{\bm{\varepsilon}}^{D-1}_{t},\widehat{\varepsilon}_{t+\tau})$ out-of-sample &
  34.26 &
  28.13 &
  69.49 &
  22.75 &
  47.05 &
  121.88 \\
$\text{K}(\bm{x}^{D-1}_{t},\widehat{\varepsilon}_{t+\tau})$ out-of-sample &
  33.11 &
  22.88 &
  41.44 &
  27.74 &
  70.20 &
  120.70 \\
PE sufficiency test &
  {\color[HTML]{3166FF} Accept} &
  {\color[HTML]{3166FF} Accept} &
  Inconclusive &
  {\color[HTML]{3166FF} Accept} &
  Inconclusive &
  Inconclusive \\
BDS test &
  Accept &
  Accept &
  Reject &
  Reject &
  Reject &
  Reject \\ \midrule
\multicolumn{1}{c}{GARCH} &
   &
   &
   &
   &
   &
   \\ \midrule
prediction error rate&
  \cellcolor[HTML]{DAE8FC}1.37\% &
  \cellcolor[HTML]{DAE8FC}0.48\% &
  \cellcolor[HTML]{DAE8FC}0.59\% &
  \cellcolor[HTML]{FE996B}34.73\% &
  \cellcolor[HTML]{FE996B}18.34\% &
  \cellcolor[HTML]{FE996B}16.49\% \\
adjMSEratio &
  92.98\% &
  89.26\% &
  95.31\% &
  83.62\% &
  78.73\% &
  80.16\% \\
bias ratio &
  0.00\% &
  0.11\% &
  0.00\% &
  0.00\% &
  0.04\% &
  0.01\%  \\
$\text{K}(\widehat{\bm{\varepsilon}}^{D-1}_{t},\widehat{\varepsilon}_{t+\tau})$ in-sample &
  22.12 &
  25.86 &
  308.66 &
  77.40 &
  128.25 &
  127.59 \\
$\text{K}(\bm{x}^{D-1}_{t},\widehat{\varepsilon}_{t+\tau})$ in-sample &
  27.47 &
  25.85 &
  183.12 &
  110.17 &
  154.34 &
  148.01 \\
PE sufficiency test &
  {\color[HTML]{3166FF} Accept} &
  {\color[HTML]{3166FF} Accept} &
  Inconclusive &
  {\color[HTML]{FE0000} Reject} &
  Inconclusive &
  Inconclusive \\
$\text{K}(\widehat{\bm{\varepsilon}}^{D-1}_{t},\widehat{\varepsilon}_{t+\tau})$ out-of-sample &
  33.44 &
  28.36 &
  69.80 &
  33.82 &
  43.24 &
  112.87 \\
$\text{K}(\bm{x}^{D-1}_{t},\widehat{\varepsilon}_{t+\tau})$ out-of-sample &
  32.07 &
  22.31 &
  44.07 &
  34.83 &
  64.29 &
  105.93 \\
PE sufficiency test &
  {\color[HTML]{3166FF} Accept} &
  {\color[HTML]{3166FF} Accept} &
  Inconclusive &
  {\color[HTML]{3166FF} Accept} &
  {\color[HTML]{FE0000} Reject} &
  Inconclusive \\
BDS test &
  Accept &
  Accept &
  Reject &
  Reject &
  Reject &
  Reject \\ \midrule
\multicolumn{1}{c}{GP} &
   &
   &
   &
   &
   &
   \\ \midrule
$d$ &
  9 &
  8 &
  12 &
  4 &
  1 &
  1 \\
prediction error rate&
  \cellcolor[HTML]{DAE8FC}2.62\% &
  \cellcolor[HTML]{FFFFC7}11.09\% &
  \cellcolor[HTML]{FE996B}73.26\% &
  \cellcolor[HTML]{FFFFC7}8.24\% &
  \cellcolor[HTML]{DAE8FC}2.66\% &
  \cellcolor[HTML]{DAE8FC}1.43\% \\
adjMSEratio &
  93.12\% &
  90.95\% &
  105.29\% &
  75.94\% &
  75.49\% &
  73.69\% \\
bias ratio &
  0.00\% &
  0.20\% &
  0.00\% &
  0.01\% &
  0.11\% &
  0.00\%  \\
$\text{K}(\widehat{\bm{\varepsilon}}^{D-1}_{t},\widehat{\varepsilon}_{t+\tau})$ in-sample &
  19.77 &
  24.23 &
  114.51 &
  37.35 &
  23.98 &
  361.2 \\
$\text{K}(\bm{x}^{D-1}_{t},\widehat{\varepsilon}_{t+\tau})$ in-sample&
  18.76 &
  28.84 &
  162.46 &
  25.39 &
  21.66 &
  87.77 \\
PE sufficiency test &
  {\color[HTML]{3166FF} Accept} &
  {\color[HTML]{3166FF} Accept} &
  {\color[HTML]{FE0000} Reject} &
  {\color[HTML]{3166FF} Accept} &
  {\color[HTML]{3166FF} Accept} &
  Inconclusive \\
$\text{K}(\widehat{\bm{\varepsilon}}^{D-1}_{t},\widehat{\varepsilon}_{t+\tau})$ out-of-sample &
  31.56 &
  31.77 &
  44.08 &
  18.09 &
  17.77 &
  97.37 \\
$\text{K}(\bm{x}^{D-1}_{t},\widehat{\varepsilon}_{t+\tau})$ out-of-sample &
  32.82 &
  23.44 &
  48.32 &
  21.14 &
  20.14 &
  41.49 \\
PE sufficiency test &
  {\color[HTML]{3166FF} Accept} &
  {\color[HTML]{3166FF} Accept} &
  Inconclusive &
  {\color[HTML]{3166FF} Accept} &
  {\color[HTML]{3166FF} Accept} &
  Inconclusive \\
BDS test &
  Accept &
  Accept &
  Reject &
  Reject &
  Reject &
  Reject \\ \midrule
\multicolumn{1}{c}{SVR} &
   &
   &
   &
   &
   &
   \\ \midrule
$d$ &
  8 &
  8 &
  8 &
  4 &
  1 &
  1 \\
prediction error rate&
  \cellcolor[HTML]{DAE8FC}4\% &
  \cellcolor[HTML]{DAE8FC}1.26\% &
  \cellcolor[HTML]{FE996B}43.12\% &
  \cellcolor[HTML]{FFFFC7}11.58\% &
  \cellcolor[HTML]{DAE8FC}0.25\% &
  \cellcolor[HTML]{FFFFC7}5.35\% \\
adjMSEratio &
  93.04\% &
  89.38\% &
  96.51\% &
  76.87\% &
  74.51\% &
  74.15\% \\
bias ratio &
  0.00\% &
  2.95\% &
 10.58\% &
  0.00\% &
  4.44\% &
  6.48\%  \\
$\text{K}(\widehat{\bm{\varepsilon}}^{D-1}_{t},\widehat{\varepsilon}_{t+\tau})$ in-sample &
  22.92 &
  24.24 &
  156.90 &
  29.36 &
  29.87 &
  203.73 \\
$\text{K}(\bm{x}^{D-1}_{t},\widehat{\varepsilon}_{t+\tau})$ in-sample &
  22.97 &
  26.88 &
  135.99 &
  28.84 &
  19.43 &
  32.20 \\
PE sufficiency test &
  {\color[HTML]{3166FF} Accept} &
  {\color[HTML]{3166FF} Accept} &
  Inconclusive &
  {\color[HTML]{3166FF} Accept} &
  {\color[HTML]{3166FF} Accept} &
  {\color[HTML]{3166FF} Accept} \\
$\text{K}(\widehat{\bm{\varepsilon}}^{D-1}_{t},\widehat{\varepsilon}_{t+\tau})$ out-of-sample &
  39.12 &
  27.43 &
  43.51 &
  22.43 &
  20.74 &
  80.20 \\
$\text{K}(\bm{x}^{D-1}_{t},\widehat{\varepsilon}_{t+\tau})$ out-of-sample &
  38.13 &
  22.56 &
  51.87 &
  31.19 &
  23.45 &
  50.23 \\
PE sufficiency test&
  {\color[HTML]{3166FF} Accept} &
  {\color[HTML]{3166FF} Accept} &
  Inconclusive &
  {\color[HTML]{3166FF} Accept} &
  {\color[HTML]{3166FF} Accept} &
 Inconclusive \\
BDS test &
  Accept &
  Accept &
  Reject &
  Reject &
  Reject &
  Reject \\ \bottomrule
\label{tb:KKYtestsimulation}
\end{longtable}
\end{footnotesize} 

In addition, Table~\ref{tb:KKYtestsimulation} also provides several forecasting performance metrics that do not require knowledge of the underlying data generating process and thus can be computed in the empirical analysis. We split the widely-used prediction accuracy metric mean squared error (MSE) into two components, namely the adjMSEratio and the bias ratio. The adjMSEratio and bias ratio are specified as follows
\begin{align*}
\text{adjMSEratio}&=\frac{\frac{1}{N}\sum^{N}_{t=1}\left[\left(x_{t}-\widehat{x}_{t}\right)-\frac{1}{N}\sum^{N}_{t=1}\left(x_{t}-\widehat{x}_{t}\right)\right]^{2}}{\text{var}(x_{t})}, \\
\text{bias ratio}&=\frac{\left[\frac{1}{N}\sum^{N}_{t=1}\left(x_{t}-\widehat{x}_{t}\right)\right]^2}{\text{var}(x_{t})}.
\end{align*} 
The former indicates how well the employed model replicates the deterministic function governing the investigated data, whereas the latter accounts for the systematic bias in estimating the constant term. The denominator, namely $\text{var}(x_{t})$, is included to make the statistics more informative. The metric adjMSEratio indicates the proportion of unexplained variation in the investigated data left by the postulate predictor. Similarly, the metric bias ratio reflects the systematic bias introduced by the considered model relative to the investigated data variance. 

The PE model sufficiency test is based on the value of statistics $\text{K}(\widehat{\bm{\varepsilon}}^{D-1}_{t},\widehat{\varepsilon}_{t+\tau})$ and $\text{K}(\bm{x}^{D-1}_{t},\widehat{\varepsilon}_{t+\tau})$. Table~\ref{tb:KKYtestsimulation} records the value of $\text{K}(\widehat{\bm{\varepsilon}}^{D-1}_{t},\widehat{\varepsilon}_{t+\tau})$ and $\text{K}(\bm{x}^{D-1}_{t},\widehat{\varepsilon}_{t+\tau})$ of the competing models fitted to each simulated series with segment length $D=4$ and delay $\tau=1$. Using Monte Carlo simulations, we construct an estimate of the 95\%   
C.I. of $\text{K}(\widehat{\bm{\varepsilon}}^{D-1}_{t},\widehat{\varepsilon}_{t+\tau})$ [$\text{K}(\bm{x}^{D-1}_{t},\widehat{\varepsilon}_{t+\tau})$] under independence. The simulation indicates the 95\%   
C.I. of $\text{K}(\widehat{\bm{\varepsilon}}^{D-1}_{t},\widehat{\varepsilon}_{t+\tau})$ [$\text{K}(\bm{x}^{D-1}_{t},\widehat{\varepsilon}_{t+\tau})$] with $D=4$ under independence is around 40. Therefore, if  
\begin{align*}
\text{K}(\widehat{\bm{\varepsilon}}^{D-1}_{t},\widehat{\varepsilon}_{t+1})<40 \quad \text{and} \quad \text{K}(\bm{x}^{D-1}_{t},\widehat{\varepsilon}_{t+1})<40, 
\end{align*}
we conclude that the point forecast is sufficient. When $\text{K}(\bm{x}^{D-1}_{t},\widehat{\varepsilon}_{t+1})$ is greater than 40, we use the block bootstrapping method to estimate a critical value of $\text{K}(\bm{x}^{D-1}_{t},\widehat{\varepsilon}_{t+1})-\text{K}(\widehat{\bm{\varepsilon}}^{D-1}_{t},\widehat{\varepsilon}_{t+1})$ to determine whether $\text{K}(\bm{x}^{D-1}_{t},\widehat{\varepsilon}_{t+1})$ is significantly greater than $\text{K}(\widehat{\bm{\varepsilon}}^{D-1}_{t},\widehat{\varepsilon}_{t+1})$. If that is the case, we reject the null hypothesis and conclude the point forecast is inadequate. Elsewhere, no affirmative conclusions can be drawn. To contrast our proposed test with the BDS test, the inferences of the BDS test are also included in the table as a comparison.

According to the results provided in Table~\ref{tb:KKYtestsimulation}, several conclusions can be drawn. Firstly, around 80\% of the time, the PE model sufficiency test can successfully indicate the sufficiency or insufficiency of the employed model. Around 20\% of the time, we find that no confirmatory inferences can be drawn. The proposed test can be inconclusive when the dependence relation within innovations is too strong and excessively overweights the residual deterministic relation $g^{r}(\cdot)$ left by the predictor. Second, by contrasting the inferences made from the conventional BDS test with that from our proposed test. As we expected, the BDS test conclusions are valid when the additive innovations of the underlying dynamics are iid. However, when the additive innovations are non-white, the BDS test can erroneously reject the model that generates the oracle point forecast.    

Additionally, by comparing the in-sample and out-of-sample inferences drawn from the PE model sufficiency test, the result suggests that the capability of the sufficiency test in distinguishing sufficient and inadequate predicting models increases with data size. The simulation shows that as the data size increases, the discrepancy between $\text{K}(\widehat{\bm{\varepsilon}}^{D-1}_{t},\widehat{\varepsilon}_{t+\tau})$ and $\text{K}(\bm{x}^{D-1}_{t},\widehat{\varepsilon}_{t+\tau})$ widens when they capture a different level of dependence structure. Take the GP predictor on X3 as an example, $\text{K}(\bm{x}^{D-1}_{t},\widehat{\varepsilon}_{t+\tau})$ computed on the out-of-sample residuals only marginally exceeds $\text{K}(\widehat{\bm{\varepsilon}}^{D-1}_{t},\widehat{\varepsilon}_{t+\tau})$. Based on the acceptance/rejection rule provided in \eqref{eqn:KKYA} and \eqref{eqn:KKYB}, we are unable to make the affirmative inference of the GP predictor. However, in the in-sample test, with $\text{K}(\bm{x}^{D-1}_{t},\widehat{\varepsilon}_{t+\tau})$ being significantly greater than $\text{K}(\widehat{\bm{\varepsilon}}^{D-1}_{t},\widehat{\varepsilon}_{t+\tau})$, our test can make an affirmative reject inference.

It is worth noting that an exception seems to occur at the oracle section for $X3$. Table 3 shows an \enquote{inconclusive} decision for X3 based on in-sample residuals and an \enquote{Accept} decision on out-of-sample residuals. The result seems to contradict the statement. However, in the case of the simulated series $\text{X3}$ and when the predictor is the oracle, the \enquote{inconclusive} inference is an example where our test cannot tell whether the predictor is sufficient or not. The \enquote{inconclusive} decision is caused by the fact that the innovations have strong dependent structures, which also leads to a significant correlation between innovation and lagged observations. In this sense, \enquote{inconclusive} is the \enquote{right} inference. With that being said, the \enquote{accept} inference from the in-sample data set is also not \enquote{wrong} as our test statistics suggest the dependence between innovation and the lagged observations are insignificant. This is evidence that the predictor captures most of the temporal structures within the observed time series.

In terms of the point forecast ability of ARMA, GARCH, GP, and SVR models, each model has its strengths and weaknesses. Linear models, ARMA and GARCH, cannot replicate nonlinear deterministic relations as evidenced by their poor performance in predicting $X4$, $X5$, and $X6$, which are governed by the nonlinear deterministic function $g(\cdot)$. However, the more complex and non-parametric model SVR can be susceptible to the asymmetric distribution of additive innovations. In the SVR model's attempt to predict simulated series with asymmetrically distributed innovations ($X2$ and $X6$), the asymmetry leads to substantial derivations in estimating the constant term. More importantly, both GP and SVR models are severely undermined by the dynamical structures present in the innovations (shown in their poor performance in predicting $X3$), especially when the investigated series has high noise contamination ratios. On the contrary, the simpler models, such as ARMA and GARCH, are robust to dynamic innovations.

\section{Modeling high-frequency foreign exchange rates} \label{sec:4}

This section applies the PE model sufficiency test to evaluate the point forecast performance of ARMA, GARCH, GP and SVR models in predicting one-hour EUR/USD rate realized volatilities. The realized volatility facilitates and improves the accuracy of the volatility measure and prediction \citep{hansen2011forecasting}. We choose to investigate the one-hour realized volatility instead of the common choice of the daily interval to ensure the adequate data size required by the PE sufficiency test. The one-hour realized volatility is computed from aggregating the squared returns of EUR/USD in every 10-minute interval. 

Letting $X=\{x_{i}; i=1,...,N+1\}$ denote the logarithm of the close bid rate of EUR/USD, the return series $\
R=\{r_{i};i=1,...,N\}$ is computed through 
\begin{align*}
r_{i}=x_{i}-x_{i-1},
\end{align*} 
and the squared returns are denoted by $%
\text{Rsq}=\{r_{i}^{2};i=1,...,N\}$.
The close bid price of EUR/USD exchange rate is provided by Thomson Reuters Tick History (\url{https://www.refinitiv.com/en/financial-data/market-data/tick-history}). The one-hour realized volatility is obtained using
\begin{equation*}
\text{rv}_{t}=\sum_{i=1+6(t-1)}^{6t}{r_{i}^2}, \quad t=1,2,\ldots, N/6.
\end{equation*}
We choose to compile data over a six-year period and divide the data set into six non-overlapping one-year periods to compare the empirical results from different investigation periods. The EUR/USD close bid is recorded
from 21:00 GMT 16/06/2013 to 20:50 GMT 21/06/2019, with the weekend entries
removed(from 21:00 GMT Friday to 20:50 GMT Sunday inclusive). Each data period starts from the 3\textsuperscript{rd} Sunday of June ends on the 3\textsuperscript{rd} Friday of June next year. There are 6360 entries in each one-year-long realized volatility series, and we have six of them, which corresponds to different years' of data. 

\begin{figure}[htbp]
\centering
\subfloat[]{
\includegraphics[width=8cm]{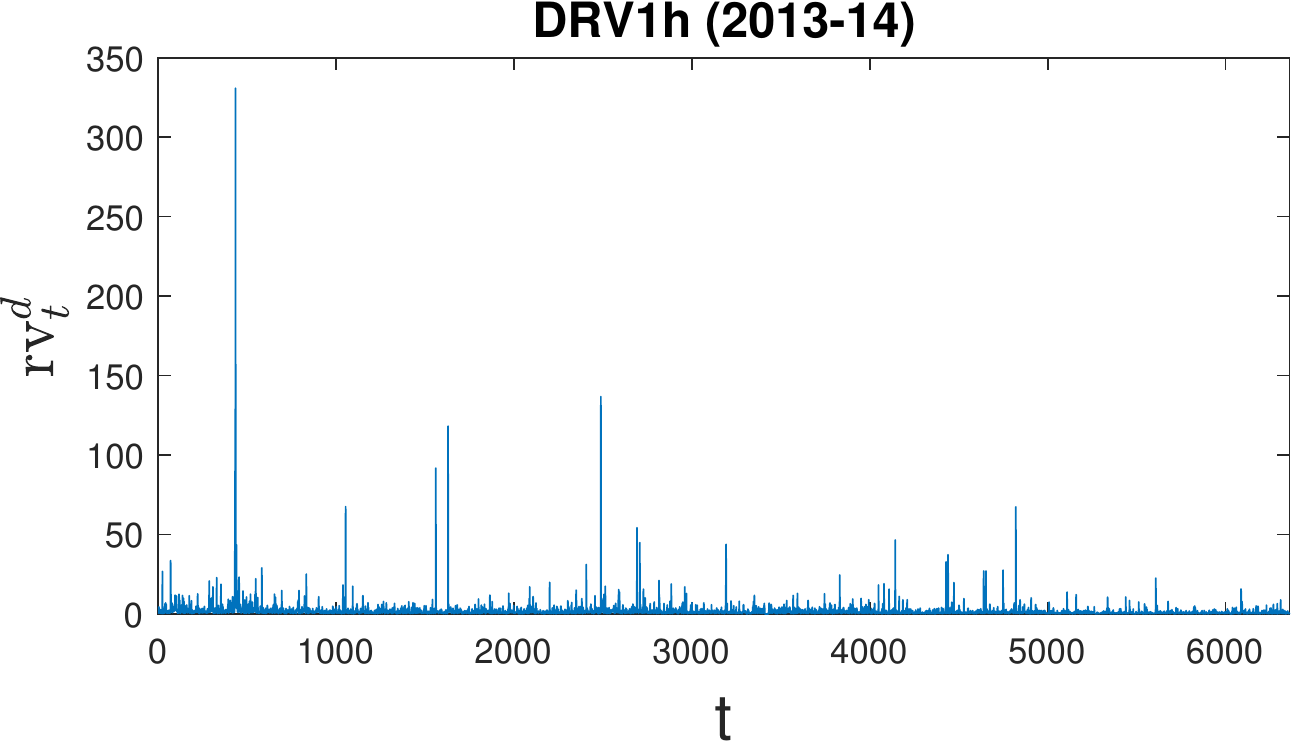}
\label{fig3:subfig1}}
\hfil
\subfloat[]{
\includegraphics[width=8cm]{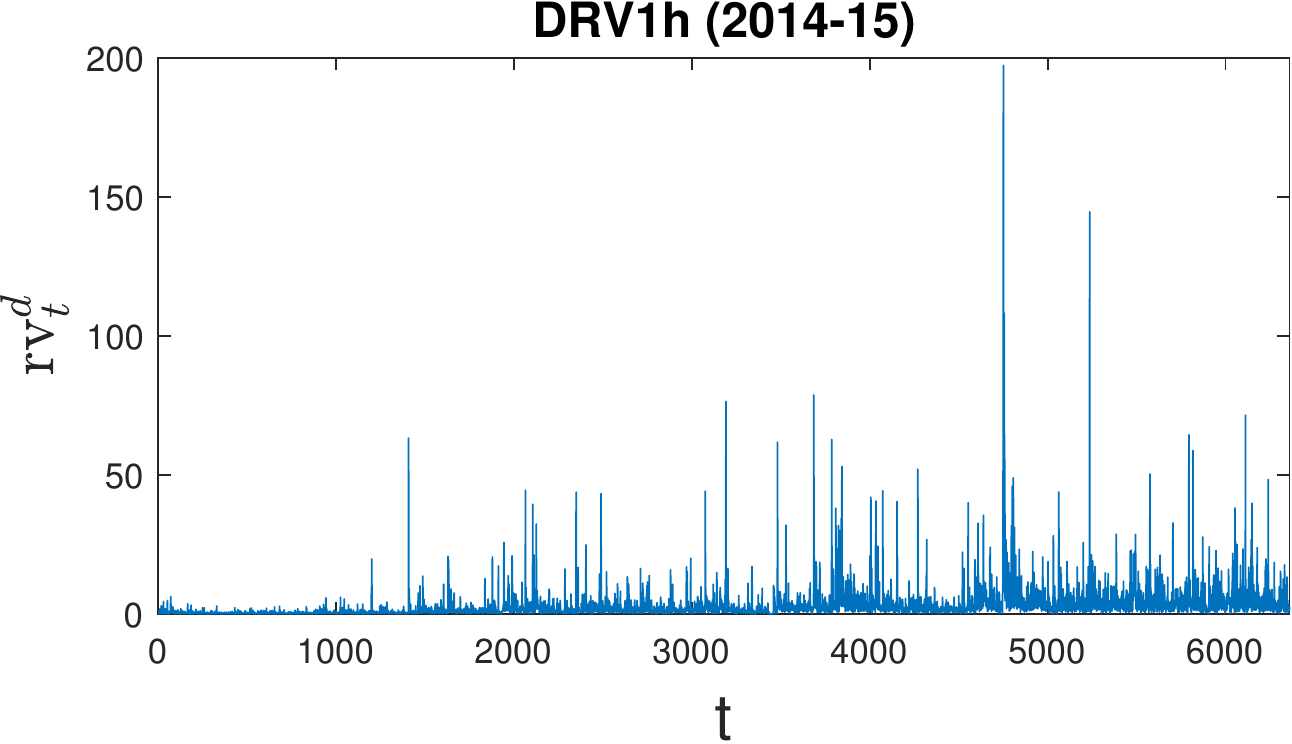}
\label{fig3:subfig4}}

\subfloat[]{
\includegraphics[width=8cm]{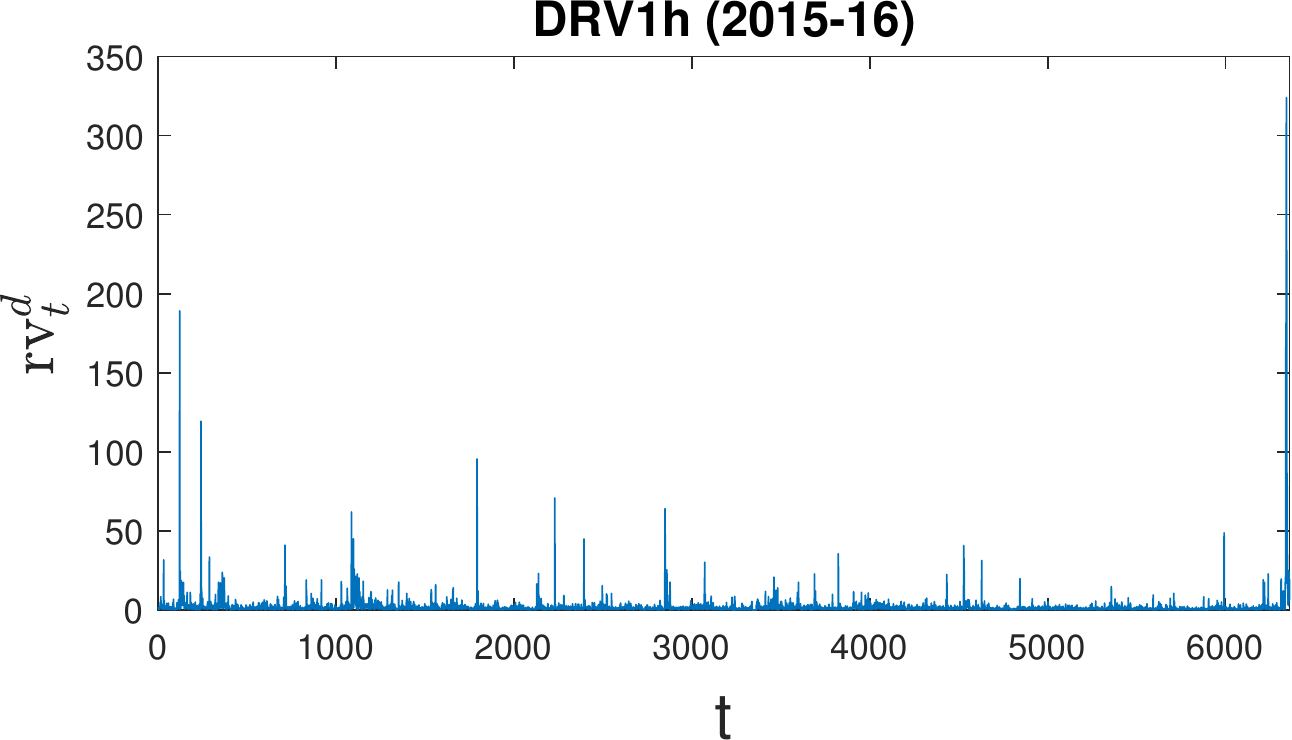}
\label{fig3:subfig2}}
\hfil
\subfloat[]{
\includegraphics[width=8cm]{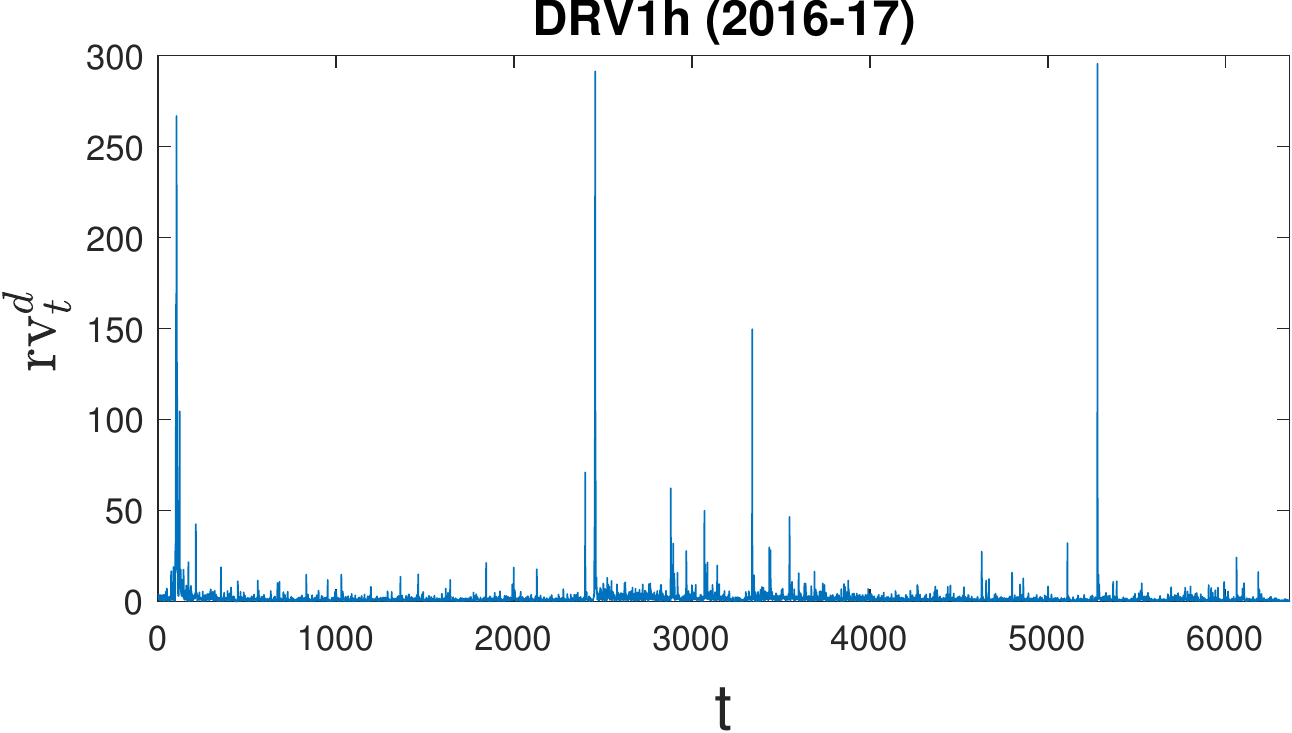}
\label{fig3:subfig5}}

\subfloat[]{
\includegraphics[width=8cm]{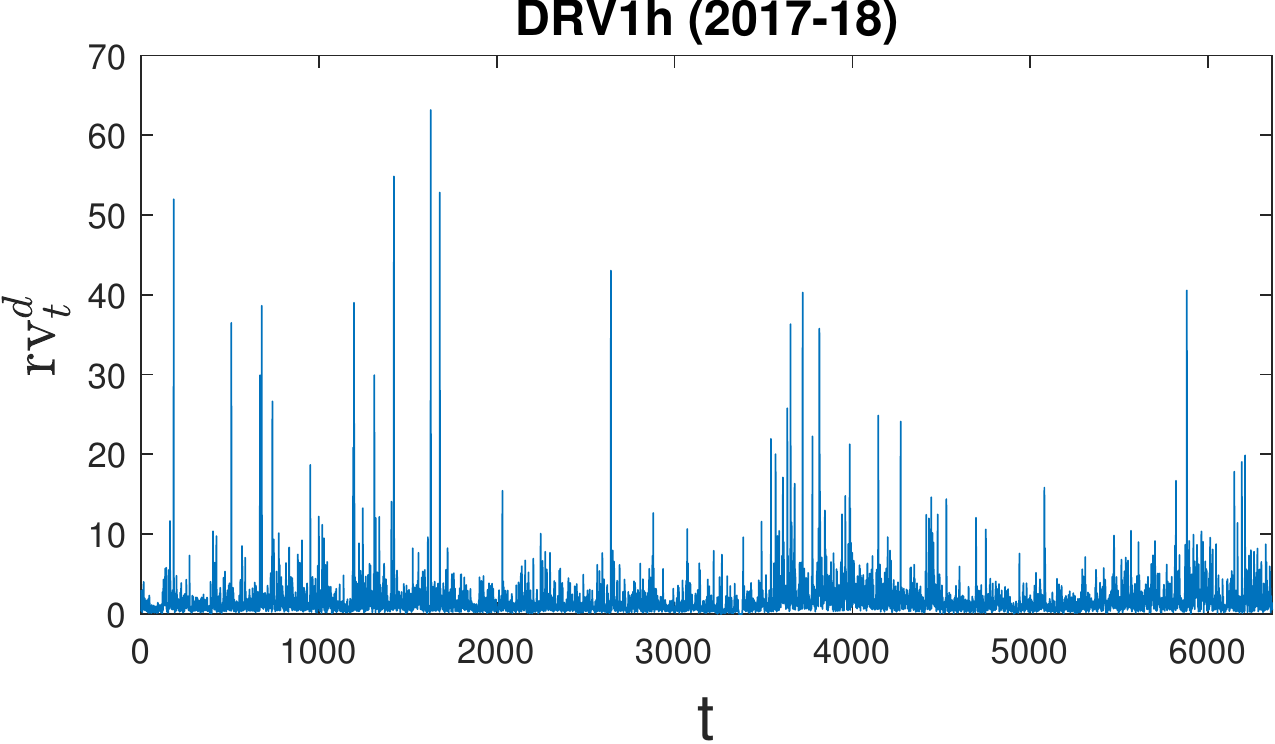}
\label{fig3:subfig3}}
\hfil
\subfloat[]{
\includegraphics[width=8cm]{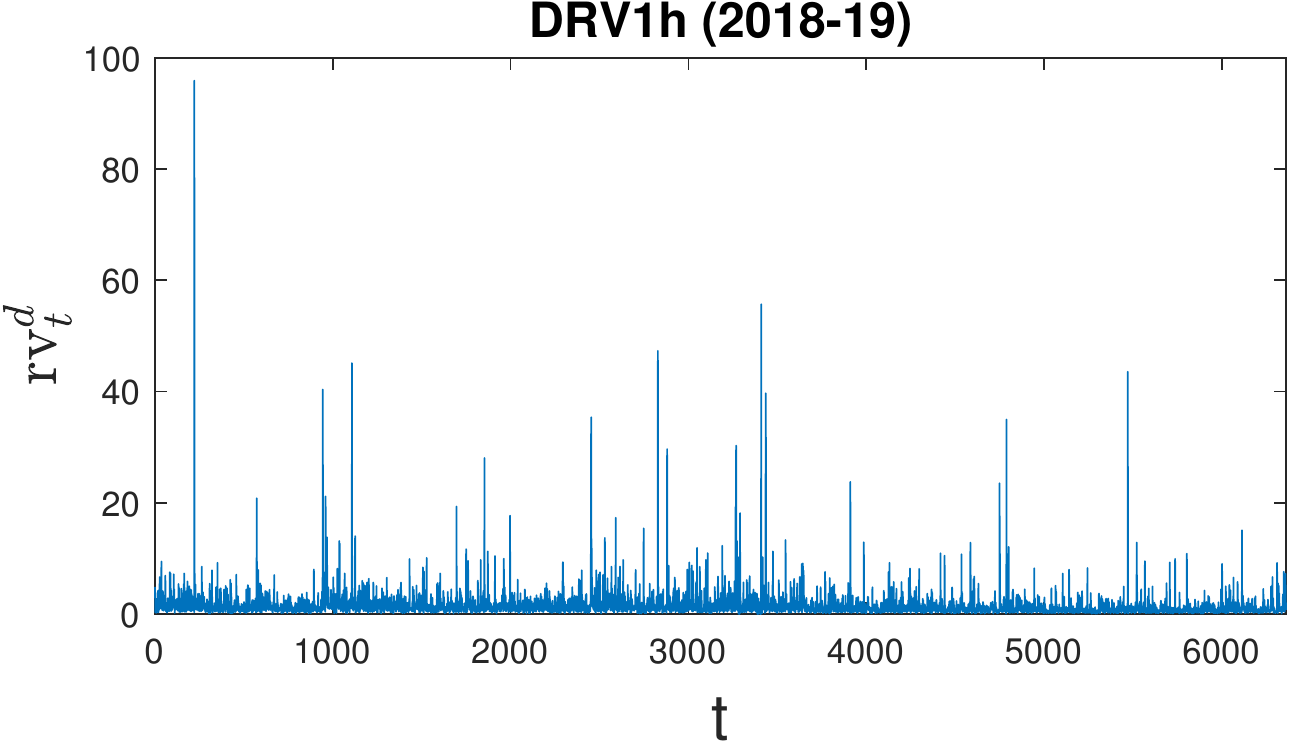}
\label{fig3:subfig6}}
\caption{Plot of six non-overlapping one-year-long  deseasonalized EUR/USD one-hour realized volatility series.}
\label{fig3}
\end{figure}

Before we fit the realized volatility time series to the prediction models, we employ \citeauthor{andersen1998deutsche}'s \citeyearpar{andersen1998deutsche} Flexible Fourier Form (FFF) method to deseasonalize it to remove the intraday periodicity. Intraday periodicity in volatility is a commonly observed feature in intraday financial time series \citep{martens2002comparison}. The periodicity in intraday foreign exchange rate returns is a 24-hour pattern which is mainly attributed to the differences in trading times in the global foreign exchange markets. The deseasonalized one-hour realized volatility is denoted by $\text{DRV1h}=\{\text{rv}^{d}_{t};t=1,2,\ldots,6360\}$ and are plotted in Figure~\ref{fig3}. 

The models used for prediction are the same models employed in the simulation studies, namely the ARMA, GARCH, GP and SVR models. Similar to the prediction procedures followed in the simulation studies, for parametric models, the first 5160 observations (around 80\% of total length) are used to estimate the employed models' parameters. The last 1200 observations are used in assessing out-of-sample prediction accuracy. For non-parametric models, the first 4000 observations are used to train the prediction model. The subsequent 1160 observations are used to compare and select the optimal hyperparameters and determine the number of lagged observations $d$ used as input variables. The last 1200 observations are used in assessing out-of-sample prediction accuracies. 

We apply the PE model sufficiency test to assess the sufficiency of the point forecast produced by the considered models. Table~\ref{tab:DRV1hPrediction} records the inference drawn from the PE sufficiency test and the value of the test statistics. Additionally, each model's estimated parameters and its out-of-sample prediction accuracy metrics, namely the adjMSEratio and the bias ratio, introduced in the previous section, are also included.  
\begin{table}[htbp]
\centering
\caption{{\footnotesize Comparison of prediction performances of ARMA, GARCH, GP and SVR models on deseasonalized EUR/USD one-hour realized volatilities in six non-overlapping one-year investigation periods. $\text{MSE}/\sum_{t}(x_{t})^{2}$ measures the mean squared error of the prediction relative to the overall variations of the investigated data. adjMSEratio and bias ratio measure how well the predictor replicates the deterministic function $g(\cdot)$ governing the investigated data and the error in estimating the constant term $c$: $\text{adjMSEratio}=\frac{1}{N}\sum^{N}_{t=1}\left[(x_{t}-\widehat{x}_{t})-\frac{1}{N}\sum^{N}_{t=1}(x_{t}-\widehat{x}_{t})\right]^{2}/\text{var}(x_{t})$; $\text{bias ratio}=\left[\frac{1}{N}\sum^{N}_{t=1}(x_{t}-\widehat{x}_{t})\right]^2/\text{var}(x_{t})$. $\text{K}(\widehat{\bm{\varepsilon}}^{D-1}_{t},\widehat{\varepsilon}_{t+\tau})$ and $\text{K}(\bm{x}^{D-1}_{t},\widehat{\varepsilon}_{t+\tau})$ with $D=4$ and $\tau=1$ are the statistics required in the PE model sufficient test. They are computed in the training set due to the inadequate data length of the investigated data in the test set. The cell filled with orange color corresponds to the models that are concluded insufficient. The model with smallest out-of-sample adjMSEratio is highlighted with red color, and is termed as \enquote{Function Estimation Winner}. The model with the smallest out-of-sample $\text{MSE}/\sum_{t}(x_{t})^{2}$ is termed as \enquote{Overall Winner}.}}
\resizebox{\textwidth}{!}{%
\begin{tabular}{@{}lllllll@{}}
\toprule
Overall Winner &
  SVR &
  ARMA &
  GARCH &
  SVR &
  ARMA &
  SVR \\ \midrule
Function Estimation Winner &
  GARCH &
  SVR &
  GARCH &
  SVR &
  SVR &
  SVR \\ \midrule
 &
  DRV1h2013-14 &
  DRV1h2014-15 &
  DRV1h2015-16 &
  DRV1h2016-17 &
  DRV1h2017-18 &
  DRV1h2018-19 \\ \midrule
\multicolumn{1}{c}{ARMA} &
   &
   &
   &
   &
   &
   \\ \midrule
$\widehat{\phi}_{1}$,$\widehat{\theta}_{1}$ &
  \cellcolor[HTML]{FFCE93}0.79, 0.63 &
  \cellcolor[HTML]{FFCE93}0.76, 0.55 &
  \cellcolor[HTML]{FFCE93}0.87, 0.75 &
  \cellcolor[HTML]{FFCE93}0.86, 0.41 &
  \cellcolor[HTML]{FFCE93}0.88, 0.77 &
  \cellcolor[HTML]{FFCE93}0.57, 0.39 \\
$\text{MSE}/\sum_{t}(x_{t})^{2}$ &
  \cellcolor[HTML]{FFCE93}81.70\% &
  \cellcolor[HTML]{FFCE93}69.03\% &
  \cellcolor[HTML]{FFCE93}63.71\% &
  \cellcolor[HTML]{FFCE93}117.85\% &
  \cellcolor[HTML]{FFCE93}56.40\% &
  \cellcolor[HTML]{FFCE93}78.92\% \\
adjMSEratio &
  \cellcolor[HTML]{FFCE93}{94.53\%} &
  \cellcolor[HTML]{FFCE93}{100.95\%} &
  \cellcolor[HTML]{FFCE93}{65.17\%} &
  \cellcolor[HTML]{FFCE93}{120.61\%} &
  \cellcolor[HTML]{FFCE93}{96.64\%} &
  \cellcolor[HTML]{FFCE93}{99.11\%} \\
bias ratio &
  \cellcolor[HTML]{FFCE93}18.11\% &
  \cellcolor[HTML]{FFCE93}0.50\% &
  \cellcolor[HTML]{FFCE93}0.07\% &
  \cellcolor[HTML]{FFCE93}0.04\% &
  \cellcolor[HTML]{FFCE93}0.23\% &
  \cellcolor[HTML]{FFCE93}5.91\% \\
$\text{K}(\widehat{\bm{\varepsilon}}^{D-1}_{t},\widehat{\varepsilon}_{t+\tau})$ &
  \cellcolor[HTML]{FFCE93}161.44 &
  \cellcolor[HTML]{FFCE93}\textbf{104.37} &
  \cellcolor[HTML]{FFCE93}123.46 &
  \cellcolor[HTML]{FFCE93}\textbf{222.50} &
  \cellcolor[HTML]{FFCE93}119.96 &
  \cellcolor[HTML]{FFCE93}122.23 \\
$\text{K}(\bm{x}^{D-1}_{t},\widehat{\varepsilon}_{t+\tau})$ &
  \cellcolor[HTML]{FFCE93}107.04 &
  \cellcolor[HTML]{FFCE93}\textbf{150.36} &
  \cellcolor[HTML]{FFCE93}119.60 &
  \cellcolor[HTML]{FFCE93}\textbf{289.53} &
  \cellcolor[HTML]{FFCE93}103.21 &
  \cellcolor[HTML]{FFCE93}121.19 \\
PE sufficiency test &
  \cellcolor[HTML]{FFCE93}Inconclusive &
  \cellcolor[HTML]{FFCE93}{\color[HTML]{FF0000} \textbf{Reject}} &
  \cellcolor[HTML]{FFCE93}Inconclusive &
  \cellcolor[HTML]{FFCE93}{\color[HTML]{FF0000} \textbf{Reject}} &
  \cellcolor[HTML]{FFCE93}Inconclusive &
  \cellcolor[HTML]{FFCE93}Inconclusive \\ \midrule
\multicolumn{1}{c}{GARCH} &
   &
   &
   &
   &
   &
   \\ \midrule
$\widehat{\alpha}_{1}$,$\widehat{\beta}_{1}$ &
  0.23, 0.43 &
  \cellcolor[HTML]{FFCE93}0.26, 0.55 &
  0.16, 0.74 &
  \cellcolor[HTML]{FFCE93}0.35, 0.49 &
  \cellcolor[HTML]{FFCE93}0.05, 0.94 &
  \cellcolor[HTML]{FFCE93}0.20, 0.29 \\
$\text{MSE}/\sum_{t}(x_{t})^{2}$ &
  83.13\% &
  \cellcolor[HTML]{FFCE93}70.88\% &
  56.77\% &
  \cellcolor[HTML]{FFCE93}110.81\% &
  \cellcolor[HTML]{FFCE93}56.45\% &
  \cellcolor[HTML]{FFCE93}79.14\% \\
adjMSEratio &
  {\color[HTML]{FF0000}{93.33\%}} &
  \cellcolor[HTML]{FFCE93}{103.67\%} &
  {\color[HTML]{FF0000}{58.10\%}} &
  \cellcolor[HTML]{FFCE93}{113.39\%} &
  \cellcolor[HTML]{FFCE93}{96.91\%} &
  \cellcolor[HTML]{FFCE93}{99.17\%} \\
bias ratio&
  20.55\% &
  \cellcolor[HTML]{FFCE93}0.50\% &
  0.03\% &
  \cellcolor[HTML]{FFCE93}0.05\% &
  \cellcolor[HTML]{FFCE93}0.05\% &
  \cellcolor[HTML]{FFCE93}6.12\% \\
$\text{K}(\widehat{\bm{\varepsilon}}^{D-1}_{t},\widehat{\varepsilon}_{t+\tau})$ &
  141.19 &
  \cellcolor[HTML]{FFCE93}\textbf{138.49} &
  160.25 &
  \cellcolor[HTML]{FFCE93}181.49 &
  \cellcolor[HTML]{FFCE93}223.13 &
  \cellcolor[HTML]{FFCE93}140.33 \\
$\text{K}(\bm{x}^{D-1}_{t},\widehat{\varepsilon}_{t+\tau})$ &
  129.38 &
  \cellcolor[HTML]{FFCE93}\textbf{204.21} &
  134.79 &
  \cellcolor[HTML]{FFCE93}191.99 &
  \cellcolor[HTML]{FFCE93}86.58 &
  \cellcolor[HTML]{FFCE93}144.54 \\
PE sufficiency test &
  Inconclusive &
  \cellcolor[HTML]{FFCE93}{\color[HTML]{FF0000} \textbf{Reject}} &
  Inconclusive &
  \cellcolor[HTML]{FFCE93}Inconclusive &
  \cellcolor[HTML]{FFCE93}Inconclusive &
  \cellcolor[HTML]{FFCE93}Inconclusive \\ \midrule
\multicolumn{1}{c}{GP} &
   &
   &
   &
   &
   &
   \\ \midrule
Selected lag for input &
  \cellcolor[HTML]{FFCE93}6 &
  1 &
  \cellcolor[HTML]{FFCE93}1 &
  \cellcolor[HTML]{FFCE93}3 &
  \cellcolor[HTML]{FFCE93}4 &
  \cellcolor[HTML]{FFCE93}10 \\
$\text{MSE}/\sum_{t}(x_{t})^{2}$ &
  \cellcolor[HTML]{FFCE93}84.38\% &
  69.04\% &
  \cellcolor[HTML]{FFCE93}86.62\% &
  \cellcolor[HTML]{FFCE93}98.29\% &
  \cellcolor[HTML]{FFCE93}56.48\% &
  \cellcolor[HTML]{FFCE93}81.05\% \\
adjMSEratio &
  \cellcolor[HTML]{FFCE93}{102.75\%} &
  100.68\% &
  \cellcolor[HTML]{FFCE93}{88.55\%} &
  \cellcolor[HTML]{FFCE93}{100.61\%} &
  \cellcolor[HTML]{FFCE93}{96.92\%} &
  \cellcolor[HTML]{FFCE93}{99.14\%} \\
bias ratio &
  \cellcolor[HTML]{FFCE93}13.82\% &
  0.77\% &
  \cellcolor[HTML]{FFCE93}0.14\% &
  \cellcolor[HTML]{FFCE93}0.01\% &
  \cellcolor[HTML]{FFCE93}0.08\% &
  \cellcolor[HTML]{FFCE93}8.36\% \\
$\text{K}(\widehat{\bm{\varepsilon}}^{D-1}_{t},\widehat{\varepsilon}_{t+\tau})$ &
  \cellcolor[HTML]{FFCE93}323.98 &
  366.58 &
  \cellcolor[HTML]{FFCE93}\textbf{178.35} &
  \cellcolor[HTML]{FFCE93}135.92 &
  \cellcolor[HTML]{FFCE93}132.87 &
  \cellcolor[HTML]{FFCE93}138.54 \\
$\text{K}(\bm{x}^{D-1}_{t},\widehat{\varepsilon}_{t+\tau})$ &
  \cellcolor[HTML]{FFCE93}304.35 &
  408.39 &
  \cellcolor[HTML]{FFCE93}\textbf{243.73} &
  \cellcolor[HTML]{FFCE93}126.84 &
  \cellcolor[HTML]{FFCE93}132.64 &
  \cellcolor[HTML]{FFCE93}150.56 \\
PE sufficiency test &
  \cellcolor[HTML]{FFCE93}Inconclusive &
  Inconclusive &
  \cellcolor[HTML]{FFCE93}{\color[HTML]{FF0000} \textbf{Reject}} &
  \cellcolor[HTML]{FFCE93}Inconclusive &
  \cellcolor[HTML]{FFCE93}Inconclusive &
  \cellcolor[HTML]{FFCE93}Inconclusive \\ \midrule
\multicolumn{1}{c}{SVR} &
   &
   &
   &
   &
   &
   \\ \midrule
Selected lag for input &
  \cellcolor[HTML]{FFCE93}3 &
  1 &
  \cellcolor[HTML]{FFCE93}1 &
  \cellcolor[HTML]{FFCE93}1 &
  4 &
  2 \\
$\text{MSE}/\sum_{t}(x_{t})^{2}$ &
  \cellcolor[HTML]{FFCE93}69.37\% &
  69.72\% &
  \cellcolor[HTML]{FFCE93}93.32\% &
  \cellcolor[HTML]{FFCE93}97.70\% &
  60.56\% &
  73.09\% \\
adjMSEratio &
  \cellcolor[HTML]{FFCE93}{96.24\%} &
  {\color[HTML]{FF0000} {98.59\%}} &
  \cellcolor[HTML]{FFCE93}{94.85\%} &
  \cellcolor[HTML]{FFCE93}{\color[HTML]{FE0000}{100.02\%}} &
  {\color[HTML]{FF0000}{95.54\%}} &
  {\color[HTML]{FF0000}{97.52\%}} \\
bias ratio &
  \cellcolor[HTML]{FFCE93}1.81\% &
  3.79\% &
  \cellcolor[HTML]{FFCE93}0.72\% &
  \cellcolor[HTML]{FFCE93}0.00\% &
  7.79\% &
  0.03\% \\
$\text{K}(\widehat{\bm{\varepsilon}}^{D-1}_{t},\widehat{\varepsilon}_{t+\tau})$ &
  \cellcolor[HTML]{FFCE93}73.47 &
  221.47 &
  \cellcolor[HTML]{FFCE93}152.23 &
  \cellcolor[HTML]{FFCE93}\textbf{258.70} &
  59.21 &
  83.70 \\
$\text{K}(\bm{x}^{D-1}_{t},\widehat{\varepsilon}_{t+\tau})$ &
  \cellcolor[HTML]{FFCE93}62.34 &
  272.96 &
  \cellcolor[HTML]{FFCE93}177.22 &
  \cellcolor[HTML]{FFCE93}\textbf{369.19} &
  81.11 &
  74.98 \\
PE sufficiency test &
  \cellcolor[HTML]{FFCE93}Inconclusive &
  Inconclusive &
  \cellcolor[HTML]{FFCE93}Inconclusive &
  \cellcolor[HTML]{FFCE93}{\color[HTML]{FF0000} \textbf{Reject}} &
  Inconclusive &
  Inconclusive \\ \bottomrule
\end{tabular}%
}
\label{tab:DRV1hPrediction}
\end{table}

From the table, all of the statistics $\text{K}(\widehat{\bm{\varepsilon}}^{D-1}_{t},\widehat{\varepsilon}_{t+\tau})$ and $\text{K}(\bm{x}^{D-1}_{t},\widehat{\varepsilon}_{t+\tau})$ are significant, suggesting strong temporal dependence structures exhibited in the innovation term of the realized volatility dynamics. The existence of dependence structures in the additive innovations of volatility dynamics actually well coincides with the general perception of the volatility dynamics. The GARCH model is the most well-known representation of volatility dynamics. The GARCH model assumes large volatilities are likely to be followed by large volatilities. In addition, large historical volatilities also lead to large variances of future volatilities. This second-order moment structure might be the primary source that leads to the temporal dependence within the innovation of the investigated data. 

We explained earlier in section~\ref{sec:2} that our test can be inconclusive if the dependence structure within the innovation is too strong. Because of that, our test is inconclusive around 80\% of the time in the empirical studies. However, our test is aligned with the prediction accuracy. When it finds an insufficient point predictor, any competing model with even poorer prediction accuracy is by default insufficient. In that way, the PE sufficiency test sets the benchmark of insufficiency when used in comparison studies. For instance, in the investigation period 2016-17, the SVR model is concluded insufficient by our proposed test. Based on the value of adjMSEratio, the GARCH and GP model produce poorer point forecast accuracy than the SVR model. They can also be concluded insufficient.  

It is worth noting that we choose to use the metric adjMSEratio to compare the prediction accuracy of the considered model. We choose it over the most commonly used prediction evaluation metric, MSE, as it excludes the contribution of the systematic bias introduced by the predictor. The derivations of the estimation of the constant term cannot be determined by our test. Consequently, the focus of the model evaluation study conducted in this paper is placed on whether the predictor accurately replicates the true deterministic relations underlying the investigated data.  

The models filled with orange backgrounds in Table~\ref{tab:DRV1hPrediction} are the ones that are deemed insufficient. They are either rejected by the PE sufficiency test, have worse prediction accuracy than the model rejected by the test, or are inferior to their competitors. It is worth noting that the models that are not filled with orange color are not necessarily the optimal model for predicting the investigation data. We just don't have enough evidence to reject their sufficiency. 

From Table~\ref{tab:DRV1hPrediction}, a number of conclusions can be drawn. First of all, EUR/USD one-hour realized volatilities are predictable. The adjMSEratio metric measures the proportion of unexplained variation in the overall variance of the investigated data. If no effort has been made into the prediction, then $\text{adjMSEratio}=1$. Therefore, $1-\text{adjMSEratio}=0$ gives the increment of the accuracy of the employed model versus random guess. The prediction model with the smallest adjMSEratio is considered the best performing model. Across all investigated volatility series, the increment of prediction accuracy of their respective best performing model ranges from 1.41\% to 41.9\%. Second, none of the considered models is sufficient enough to exploit the maximum prediction potential. They are all deemed insufficient in at least three investigated periods. Third, the prediction performance of the employed models varies considerably across different investigated periods. For instance, the GARCH model provides the optimal prediction accuracy compared to its competitors in predicting DRV1h in the year 2015-16. However, its prediction produces the worst prediction accuracy during the year 2014-15 and 2018-19. On the contrary, the SVR model is superior to all of its competitors in 2014-15, 2016-17, 2017-18 and 2018-19, but products particularly poor predictions in 2015-16. 

Since we found all of the ARMA, GARCH, GP and SVR models cannot fully exploit the point forecast potential of the realized volatility series under study, there must be reasons that prevent them from making accurate predictions. The GP and SVR model's suboptimal performances might be due to the dependent structures exhibited in the innovations of the investigated data. We show in the simulation study that both GP and SVR models are severely undermined by the dependent structures exhibited in the innovations, especially when the investigated series has a relatively small signal-to-noise ratio. On the contrary, the  ARMA, GARCH are robust to dependent innovations. We suspect that their poor prediction performance might be caused by non-stationary exhibited in the EUR/USD volatility dynamics. 

\begin{figure}[!t]
\centering
\includegraphics[width=\textwidth]{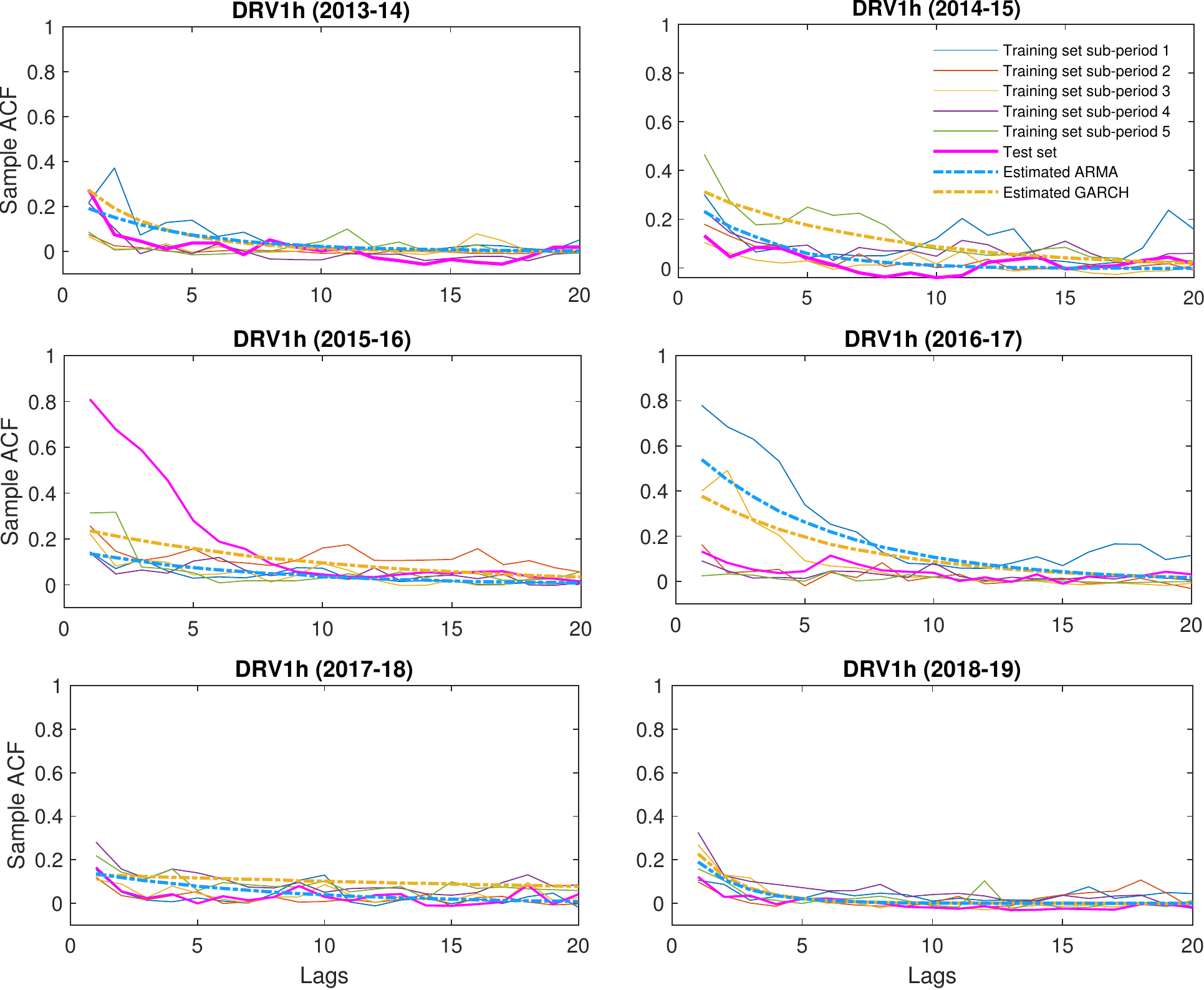}
\caption{{\small The plot of sample ACF of the deseasonalized EUR/USD one-hour realized volatilities in the six split sub-period of the original investigation period. The first five sub-periods correspond to the data used in the training set and the last sub-period in the testing set. The sample ACFs of simulations generated from the estimated ARMA(1,1) and GARCH(1,1) model are also included in conjunction to exhibit the departure of the linear deterministic structure postulated by the estimated model to that in the actual data in the testing set.}}
\label{fig4}
\end{figure}

To verify our suspicion, we split the DRV1h series into six equal-length sub-periods and compare their respective autocorrelation function (ACF). The first five sub-periods correspond to the data used in the training set and the last sub-period in the test set. The ACF of DRV1h in each sub-period is plotted in Figure~\ref{fig4}. Also, we plot the sample ACF of simulations generated from the estimated ARMA(1,1) and GARCH(1,1) model to display the departure of the linear serial relations postulated by the estimated model from that in the actual data in the test set. 

Figure~\ref{fig4} indicates that the ACF of the realized volatility series under study varies considerably over time. Not only so, but the deterministic relation governing the movements of DRV1h also seems to change abruptly in an unsystematic manner. To cope with such time-varying determinism, ARMA and GARCH model tries to match the average level of the linear serial dependence underlying the investigated data. As a result, if the average serial dependence in the training set data exceeds that in the test set, such as in the year 2016-17, the estimated ARMA and GARCH model would overestimate the deterministic structures for predicting the test set. On the contrary, when the training set exhibits weaker average serial dependence compared to that in the test set, such as in the year 2015-16, predictions made from the ARMA and GARCH model underestimate the level of determinism. In both cases, the out-of-sample point forecasts are inadequate.     

\section{Conclusions} \label{sec:5}

This paper proposed a point forecast sufficiency test that fills the conceptional incoherence between the two primary genres of model/prediction evaluation methods: the residual independence test and the prediction accuracy metrics. We show that our proposed test named the PE model sufficiency test, can remedy the limitations of mainstream model evaluation approaches. The test assesses the considered model's performance in fulfilling the point prediction task, which is the main objective for most forecasting practices.

Compared to the conventional model sufficiency test, our proposed test is not built on the assumption that a sufficient model is expected to eliminate all structures initially present in the observed data, thus resulting in independent and identically distributed residuals. We demonstrate that a sufficient point predicting model can generate non-white residuals if there exists higher than first-moment dependence relations in its underlying process. Therefore, the inferences drawn from the conventional sufficiency test, such as the well-known BDS test, can be misleading in evaluating the employed model's point prediction performance. With empirical evidence indicating the prevailing existence of higher-moment structures present in the financial return series \citep{khademalomoom2019higher}, our proposed test provides a more reliable approach in
unveiling the point forecast sufficiency of a given model in the area of financial time series analysis.  

In contrast to the prediction accuracy metrics, such as the very commonly used MSE, which aims to reflect the postulated predictors' point prediction accuracy, the PE model sufficiency test is more informative. It evaluates the performance of the considered model without the need to compare it with a benchmark. It can particularly reveal the employed model's point forecast performance relative to the maximum prediction potential of its underlying dynamic.

Our proposed test is the first attempt to assess the point forecast sufficiency of the considered models to the best of our knowledge. It is specially designed to remain valid for time series with non-white innovation when cast into an additive form. 

The simulation studies provide evidence that our newly proposed test successfully reveals the sufficiency of the point forecast performance of the employed model. We simulate a number of time series governed by different forms of data generating processes. Among the simulated time series, some exhibit linear serial correlations, and some have non-linear deterministic structures. Some are with symmetrically distributed innovations, and some have asymmetrically distributed innovations. The most important attribute of the simulated time series is the specification of the additive innovation. Some simulated series have independent innovations, and some have dependent innovations. In the simulation studies, we found the inferences drawn from the PE model sufficiency test are consistent with the \enquote{true} point forecast performance of the considered models even though it can be inconclusive if the temporal-dependent structures within innovations are too strong, which leaves room for further improvements.

The simulation studies also demonstrate the point forecast ability of ARMA, GARCH, GP, and SVR models in response to the number of potentially challenging properties commonly observed in financial time series such as non-normality, non-linearity, and dynamical innovations. The results from the simulation studies indicate that each considered model has its strengths and weaknesses. Linear models ARMA and GARCH cannot replicate non-linear deterministic relations. However, the more complex and non-parametric models SVR can be susceptible to asymmetrically distributed innovations. In particular, the asymmetrical distributions of innovation can lead to the derivation in constant estimation in the SVR model. More importantly, both GP and SVR models are severely undermined by the dependent structures exhibited in the innovations, especially when the investigated series has a relatively small signal-to-noise ratio. In contrast, simpler models, such as ARMA and GARCH, are robust. 

By applying the PE sufficiency test on the EUR/USD realized volatility series, we assess the sufficiency of ARMA, GARCH, GP and SVR models in predicting intraday foreign exchange volatilities. We found that none of the considered models is sufficient to exploit the investigated data's prediction potential. Moreover, their prediction performances vary considerably for different investigation periods. Their inadequate prediction performances are mainly due to the temporal dependence structure within innovations that undermines the GP and SVR model and the abruptly time-varying deterministic relations of the observed data that invalidates the fundamental stationary assumption required by most prediction models.        

There are several ways in which the current work may be further extended, and we briefly outline four possibilities. First, since our test can be inconclusive in certain conditions, future studies can be carried out to provide definite conclusions when our test is indecisive. Second,  due to the limited length of the paper, we mainly focused on examining our test's performance on the time series with innovations from the GARCH model where the form of dependence in the residual is caused by heteroskedasticity in the variance. The reason for our choice is that the aforementioned form of dependence is the most common structure in innovations of financial time series. Investigating our test's performance on time series with such form of dependent innovation provides a great indication of its usefulness in assessing the predicting models in financial time series. Given the length of the current paper, we didn't include simulation studies with other types of dependent innovations. One could investigate the test's performance and the behavior of the proposed test statistics on various extents of dependence relations in the generating process of innovations. This may potentially shed light on the applicability of this test in various circumstances. Third, we could derive the theoretical asymptotic distribution of the test statistics of the proposed test to obtain the $p$-value of the test. Lastly, we introduced a novel dependence measure as the test statistics in our proposed test. One may study its connection with natural measures of dependence, namely the copula, and compare it with some rank correlation statistics, such as Spearman's $\rho$ or Kendall's $\tau$.

\begin{appendices}
\section*{Appendix: The BDS test on a logarithm transformed GARCH squared standardized residuals}
\begin{figure}[!htbp]
\centering
\includegraphics[width=10.5cm]{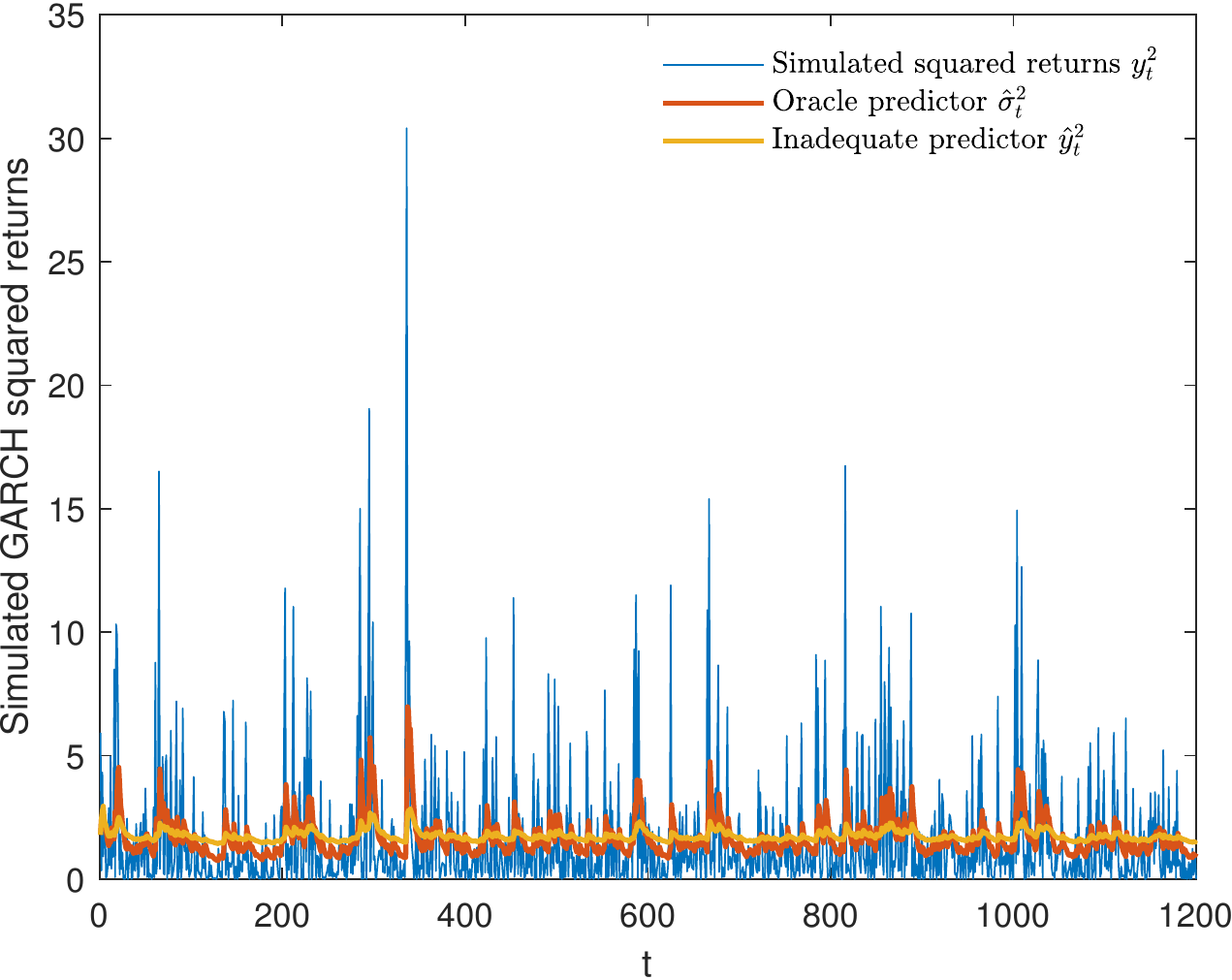}
\caption{Plot of simulated squared return time series generated from GARCH(1,1) model with parameter $\alpha_{0}=0.18$, $\alpha_{1}=0.16$, $\beta_{1}=0.74$, its oracle predictor and an inadequate predictor based on GARCH(1,1) model with parameter $\alpha_{0}=0.18$, $\alpha_{1}=0.03$, $\beta_{1}=0.87$.}
\label{fig5}
\end{figure}

The following simulation provides an example of how and why the BDS test on a logarithm transformed squared standardized residuals fails to reject an inadequate predictor/model, especially when the investigated time series has a relatively low signal-to-noise ratio. We generate a squared return time series from a GARCH(1,1) model with the parameters $\alpha_{0}=0.18$, $\alpha_{1}=0.16$ and $\beta_{1}=0.74$. In Figure~\ref{fig5}, we plot the generated time series and its oracle predictor computed from the governing formulas of the GARCH model. In addition, in the same graph, we plot an inadequate predictor based on the deterministic relation of GARCH model $\alpha_{0}=0.18$, $\alpha_{1}=0.03$ and $\beta_{1}=0.87$. The inadequate predictor underestimates the simulated target series's series correlations and is significantly different from the oracle predictor. Table~\ref{tb:BDS} provides the values of BDS statics for the nature logarithm of squared standardized residual of the inadequate predictor for varying embedding dimension $M$ from 2 to 10 of various distance $r$: $(0.25,0.5,\ldots,1.25)\times \text{s.d.}(x)$. Under the iid null hypothesis, the BDS statistic has a limiting standard normal distribution. 

\begin{table}[!hbtp]
\tabcolsep 0.3in
\centering
\caption{The BDS statistics on $\ln(y^{2}_{t}/\widehat{y}^{2}_{t})$ of embedding dimension $M$ with distance of $r$. Under null hypothesis of iid, the BDS statistic is expected to follow standard normal distribution. * indicates  the iid hypothesis is rejected at 5\% significance, and ** indicates the iid hypothesis is rejected at 1\% significance.}
\begin{tabular}{@{}lrrrrr@{}}
\toprule  
	& \multicolumn{5}{c}{$r$} \\
$M$ & $0.25\text{s.d.}(X)$ & $0.5\text{s.d.}(X)$ & $0.75\text{s.d.}(X)$ & $\text{s.d.}(X)$ & $1.25\text{s.d.}(X)$ \\ \midrule
$2$  & 0.43        & 0.24       & -0.06       & -0.26    & -0.27       \\
$3$  & -0.08       & -0.18      & -0.54       & -0.69    & -0.75       \\
$4$  & -0.34       & -0.40      & -0.66       & -0.88    & -1.00       \\
$5$  & -0.24       & -0.32      & -0.46       & -0.70    & -0.84       \\
$6$  & 0.38        & -0.38      & -0.31       & -0.58    & -0.72       \\
$7$  & -1.67       & -0.53      & -0.22       & -0.52    & -0.65       \\
$8$  & -0.46       & -0.56      & -0.15       & -0.51    & -0.64       \\
$9$  & -2.32*       & -0.61      & -0.18       & -0.51    & -0.69       \\
$10$ & -9.95**       & -0.29      & -0.34       & -0.55    & -0.73       \\ \bottomrule
\end{tabular}
\label{tb:BDS}
\end{table}

\begin{figure}[!htbp]
\centering
\subfloat[Plot of a randomly selected sub-period of the simulated GARCH squared return series $y^{2}_{t}$, and the corresponding oracle predictors $\sigma^{2}_{t}$ and inadequate predictors $\widehat{y}^{2}_{t}$.]
{\includegraphics[width=8.5cm]{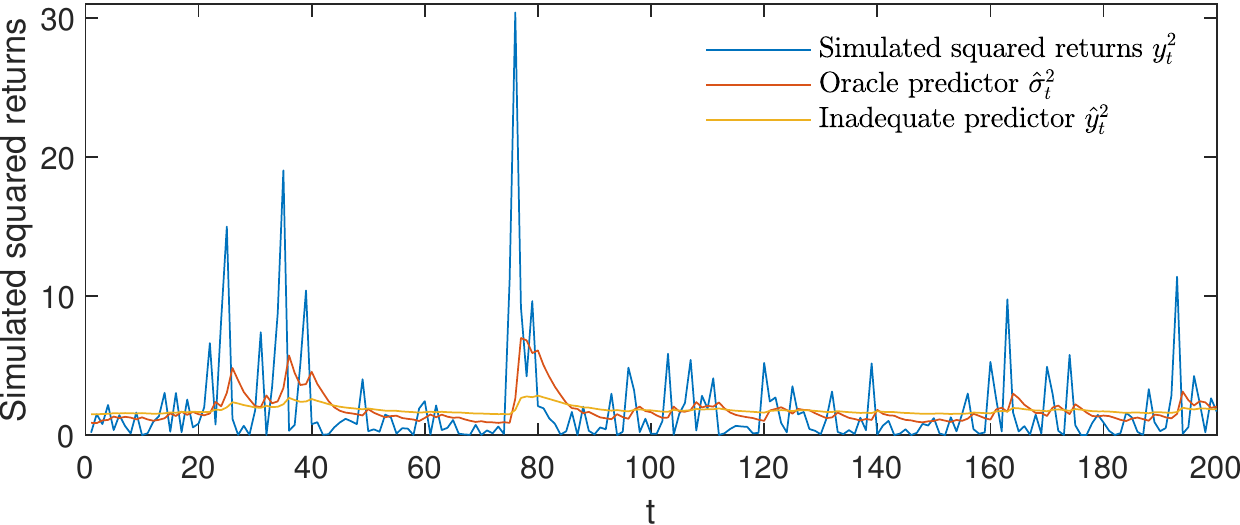}\label{fig6:subfig1}}
\quad
\subfloat[Plot of the corresponding squared standardized residuals of inadequate predictors $y^{2}_{t}/\widehat{y}^{2}_{t}$ and the logarithm of squared standardized residual of inadequate predictors $\ln(y^{2}_{t}/\widehat{y}^{2}_{t})$.]
{\includegraphics[width=8.5cm]{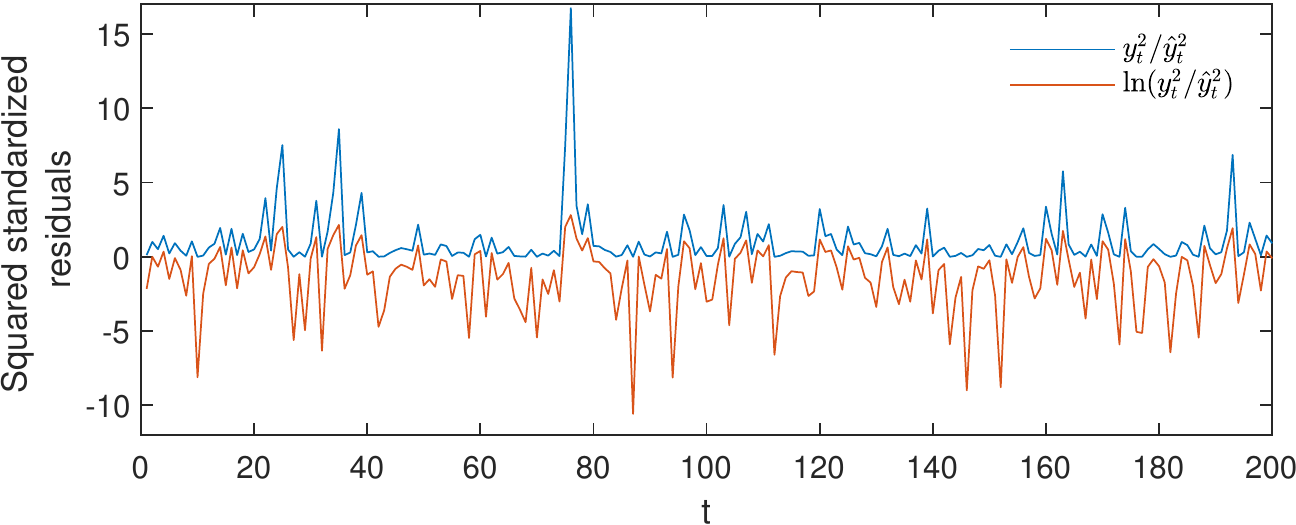}\label{fig6:subfig2}}
\caption{Effect of logarithm transformation on GARCH squared standardized residuals.}\label{fig6}
\end{figure}

From the table, except for $M=10$ and $r=0.25 \; \text{s.d.}(x)$, none of the BDS statistics is significant enough to reject the suboptimal predictors. The logarithm transformation tends to mislead the BDS test because logarithm on residuals would convert a small value close to zero to a significant negative value. Meanwhile, the logarithm transformation reduces the large valued squared standard residual where there is a distinct discrepancy between the prediction and the actual observation. As a result, the uncaptured structure becomes less evident after the logarithm transformation. Figure~\ref{fig6} compares the squared standardized residuals versus the logarithm of the squared standardized residuals of a randomly selected sub-period of the target simulation series so to display the effect of the logarithm transformation on the standardized residuals. Due to the above limitations, both the acceptance and rejection inference of the BDS test can be misleading, particularly in evaluating the performance of point prediction.

\end{appendices}

\newpage
\bibliographystyle{agsm}
\bibliography{Landy} 

\end{document}